  \documentclass[10pt,twocolumn,twoside,letterpaper]{IEEEtran}
  
  \usepackage[cmex10]{amsmath} 
  \usepackage{amsfonts}
  \usepackage{amssymb}
  \usepackage{amsthm}
  \usepackage{mathtools}
  \usepackage{cite} 
  \usepackage[dvips]{graphicx}
  \usepackage{array}
  \usepackage{color}
  \usepackage{float}
  \usepackage[caption=false,font=footnotesize]{subfig}
  \usepackage[plain]{algorithm}
  \usepackage{algpseudocode}
  \usepackage{siunitx}
  
  \graphicspath{{Pictures/}}
  \DeclareGraphicsExtensions{.eps}
  
  \restylefloat{figure} 
  
  \floatstyle{ruled} 
  \restylefloat{algorithm}
  
  \newcommand{\Bbold}{\mathbf{B}} 
  \newcommand{\Xcal}{\mathcal{X}} 
  \newcommand{\Fcal}{\mathcal{F}} 
  \newcommand{\Zcal}{\mathcal{Z}} 
  \newcommand{\Ocal}{\mathcal{O}} 
  \newcommand{\Gcal}{\mathcal{G}} 
  \newcommand{\Lcal}{\mathcal{L}} 
  \newcommand{\Nset}{\mathbb{N}} 
  \newcommand{\Rset}{\mathbb{R}} 
  \newcommand{\Prob}{\mathbb{P}} 
  \newcommand{\ds}{\displaystyle} 
  \newcommand{\Exp}{\mathbb{E}} 
  \newcommand{\var}{\mathrm{var}}

  \newtheorem{theoremint}{Theorem}
  \newtheorem{definitionint}{Definition}
  \newtheorem{exampleint}{Example}
  \newtheorem{notationint}{Notation}
  \newtheorem{propositionint}{Proposition}
  \newtheorem{corollaryint}{Corollary}
  \newtheorem{propertyint}{Property}
  \newtheorem{lemmaint}{Lemma}

    \newcommand*{\hdelimiter}{}

  \newenvironment{theorem}{\hdelimiter\begin{theoremint}}{\hdelimiter\end{theoremint}}

  \newenvironment{corollary}{\hdelimiter\begin{corollaryint}}{\hdelimiter\end{corollaryint}}
  \newenvironment{property}{\hdelimiter\begin{propertyint}}{\hdelimiter\end{propertyint}}
  \newenvironment{lemma}{\hdelimiter\begin{lemmaint}}{\hdelimiter\end{lemmaint}}
  
  \title{Regional variance for multi-object filtering}
  \author
  {
    \IEEEauthorblockN{Emmanuel Delande, Murat \"{U}ney, J\'{e}r\'{e}mie Houssineau, Daniel Clark}
    
    \thanks{Emmanuel D. Delande, J\'{e}r\'{e}mie Houssineau and Daniel E. Clark are with the School of Engineering \& Physical Sciences, Heriot-Watt University (HWU), Edinburgh (e-mails: E.D.Delande@hw.ac.uk, jh207@hw.ac.uk and d.e.clark@hw.ac.uk).}
    
    \thanks{Murat \"{U}ney is with the Institute of Digital Communications, University of Edinburgh, Edinburgh (e-mail: M.Uney@ed.ac.uk).}
    
  }

\begin{document}
  \maketitle

  \begin{abstract}
    Recent progress in multi-object filtering has led to algorithms that compute the first-order moment of multi-object distributions based on sensor measurements. The number of targets in arbitrarily selected regions can be estimated using the first-order moment. In this work, we introduce explicit formulae for the computation of the second-order statistic on the target number. The proposed concept of regional variance quantifies the level of confidence on target number estimates in arbitrary regions and facilitates information-based decisions. We provide algorithms for its computation for the Probability Hypothesis Density (PHD) and the Cardinalized Probability Hypothesis Density (CPHD) filters. We demonstrate the behaviour of the regional statistics through simulation examples.
  \end{abstract}
  
  \begin{keywords}
    Multi-object filtering, Higher-order statistics, PHD filter, CPHD filter, random finite sets, Bayesian estimation, target tracking
  \end{keywords}
  
  \section{Introduction} \label{SecIntroduction}
    Multi-target tracking dates back to the 1970s due to the requirement for aerospace or ground-based surveillance applications \cite{Bar-Shalom_Y_1978, Reid_D_1979} and involves estimating the states of a time varying number of targets using sensor measurements\cite{Blackman_SS_1999}. The Finite Set Statistics (FISST) methodology \cite{Mahler_RPS_2007_3} provides an alternative to the conventional approaches \cite{Blackman_SS_1999} in which targets are described as individual tracks, by modelling the collection of target states as a (simple) \textit{point process} or Random Finite Set (RFS). In particular, the collection of target states is a set whose size -- the number of targets -- and elements -- the states -- are both random.
    
    Multi-target RFS models lead to the well known Bayesian recursions for filtering sensor observations thereby providing a coherent Bayesian framework. These recursions, however, are not tractable for an increasing number of targets \cite{Mahler_RPS_2007_3}. Instead, the FISST methodology provides a systematic approach for approximating the Bayes optimal filtering distribution through its incomplete characterisations. Mahler's Probability Hypothesis Density (PHD) \cite{Mahler_RPS_2003} and Cardinalized Probability Hypothesis Density (CPHD) \cite{Mahler_RPS_2007} filters focus primarily on the extraction of the first moment density (also known as the intensity or the Probability Hypothesis Density) of the posterior RFS distribution, a real-valued function on the state space whose integral in any region $B$ provides the mean target number inside $B$ \cite{Mahler_RPS_2003}. A more recent filter \cite{Vo_BT_2013} has been developed in order to propagate the full posterior RFS distribution under specific assumptions on the target behaviour.
    
    In this article, we are concerned with the \textit{second-order} information on the local target number in an arbitrary region $B$, which gives a measure of uncertainty associated with the mean target number. The quantification of the confidence on the first moment density is useful for problems involved with information-based decision such as distributed sensing \cite{Uney_M_2011, Uney_M_2013, Battistelli_G_2013}, and multi-sensor estimation and control \cite{Mahler_RPS_2009, Delande_E_2011, Ristic_B_2010, Ristic_B_2011}. We propose a unified description for the first and the second-order regional statistics and derive explicit formulae for the \textit{mean target number} and the \textit{variance in target number}. The mathematical framework we introduce builds upon recent developments in multi-object modelling and filtering \cite{Clark_DE_2012_2, Clark_DE_2013_2, Houssineau_J_2013_3} and has the potential of leading to the derivations of closed form expressions for regional higher-order statistics of RFS distributions. Previous studies \cite{Mahler_RPS_2007, Singh_S_2009_2} have investigated higher-order statistics in target number, but evaluated in the whole state space and not in any arbitrary region. We provide algorithms for the computation of the regional variance using both the PHD and the CPHD filters.
    
    The structure of the article is as follows: Section \ref{SecStochasticProcesses} provides background on point processes and multi-object filtering, and introduces the regional variance in target number. In Section \ref{SecFilter}, we discuss the principles underpinning the PHD and CPHD filters before we give the details on constructing the regional statistics for the PHD and the CPHD filters, the main results of this article. In Section \ref{SecSimulation} we demonstrate the proposed concept through simulation examples and then we conclude (Section \ref{SecConclusion}). The proofs of the results in Section \ref{SecFilter} are in Appendices \ref{AppIntermediaryResults} and \ref{AppProofs}. The computational procedures are given in Appendix \ref{AppAlgorithms}.

  \section{Point processes and multi-object filtering} \label{SecStochasticProcesses}
    In this section, we introduce background and notation used throughout this article. We first give a brief review of point processes (Section \ref{SubsecDefintion}) and define the regional statistics (Section \ref{SubsecStatistics}). In Section \ref{SubsecFunctionalDifferentiation} we introduce the functional differential that is used to extract the regional statistics of point processes from their generating functionals, which are covered in Section \ref{SubsecGeneratingFunctionals}. Section \ref{SubsecBayesFiltering} overview the Bayesian framework from which the PHD and CPHD filters are constructed.
    
    \subsection{Point processes} \label{SubsecDefintion}
      In this article, the objects of interest - the \textit{targets} - have individual states $x$ in some target space $\Xcal \subset \Rset^{d_x}$, typically consisting of position and velocity variables. The multi-object filtering framework focuses on the target \textit{population} rather than individual targets. Both the target number \textit{and} the target states are unknown and (possibly) time-varying. So, we describe the target population by a point process $\Phi$ whose number of elements \textit{and} element states are random. A realisation of a point process $\Phi$ is a set of points $\varphi = \{x_1,\dots, x_{N}\}$ depicting a specific multi-target configuration.
      
      More formally, a point process $\Phi$ on $\Xcal$ is a measurable mapping:
      \begin{equation} \label{EqPointProcess}
	\Phi: (\Omega, \Fcal, \Prob) \rightarrow (E_{\Xcal}, \Bbold_{E_{\Xcal}})
      \end{equation}
      from some probability space $(\Omega, \Fcal, \Prob)$ to the measurable space $(E_{\Xcal}, \Bbold_{E_{\Xcal}})$, where $E_{\Xcal}$ is the point process state space, i.e., the space of all the finite sets of points in $\Xcal$, and $\Bbold_{E_{\Xcal}}$ is the Borel $\sigma$-algebra on $E_{\Xcal}$ \cite{Stoyan_D_1995}. We describe $\Phi$ by its probability distribution on $(E_{\Xcal}, \Bbold_{E_{\Xcal}})$ generated by $\Prob$, denoted by $P_{\Phi}$ (as in the study of random variables). The \textit{probability density} $p_{\Phi}$ of the point process $\Phi$, if it exists, is the Radon-Nikodym derivative of the probability measure $P_{\Phi}$ with respect to (w.r.t.) the Lebesgue measure.
      
      The Finite Set Statistics methodology for target tracking \cite{Mahler_RPS_2007} considers the representation of RFSs through their \textit{multi-object density} $f_{\Phi}$ (derived from $p_{\Phi}$). This approach has the distinctive merit of producing more intuitive and accessible results facilitating rather direct derivations of filtering algorithms such as the PHD filter \cite{Mahler_RPS_2003}. However, the regional variance in target number  does not necessarily admit a density, in the general case. Therefore, we chose to adopt a measure-theoretical formulation, based on more general representations of point processes \cite{Stoyan_D_1995}, \cite{VereJones_D_2008}, out of practical necessity. A thorough discussion on the relation between measures and associated densities can be found in \cite{Vo_BN_2005, Houssineau_J_2013_2}.
      
    \subsection{Regional statistics: mean and variance in target number} \label{SubsecStatistics}
      Unlike real-valued random variables, the space of point processes is not endowed with an expectation operator from which various \textit{statistical moments} could be derived. Recall from the definition \eqref{EqPointProcess} of a point process $\Phi$ that two realisations $\varphi$, $\varphi' \in E_{\Xcal}$ are sets of points. Since the sum of two sets (e.g. $\{x_1, x_2\} + \{x'_1, x'_2, x'_3\}$) is ill-defined, so would be the ``usual'' expectation operator $\Exp[\Phi]$ on point processes.
      
\begin{figure}[H]
	\centering
	\includegraphics[width=0.7\columnwidth]{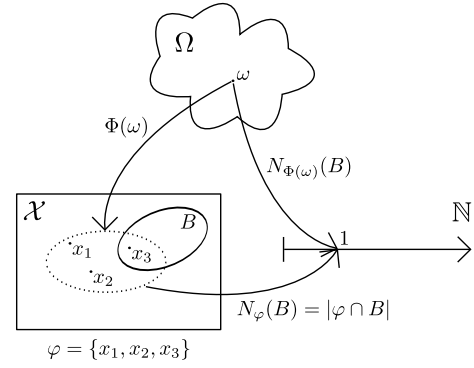}
	\caption{Point process and counting measure.\label{FigPointProcessAndCountingMeasure}} 
      \end{figure}

      Nevertheless, point processes can alternatively be described by the point patterns they produce in the target state space $\Xcal$ rather than by their realisations in the process state space $E_{\Xcal}$ (see Figure \ref{FigPointProcessAndCountingMeasure}). For any Borel set $B \in \Bbold_{\Xcal}$, where $\Bbold_{\Xcal}$ is the Borel $\sigma$-algebra on $\Xcal$, the integer-valued random variable
      \begin{equation} \label{EqCountingMeasure}
	N_{\Phi}(B) = \sum_{x \in \Phi} 1_B(x)
      \end{equation}
      counts the number of targets falling inside $B$ according to the point process $\Phi$ \cite{Stoyan_D_1995}. Using the well-defined statistical moments of the integer-valued random variables $N_{\Phi}(B)$ for any $B \in \Bbold_{\Xcal}$, one can define the \textit{moment measures} of the point process $\Phi$.   
      \noindent
      For any regions $B, B' \in \Bbold_{\Xcal}$, the first and second moment measures $\mu^{(1)}_{\Phi}$, $\mu^{(2)}_{\Phi}$ are defined by
      \begin{subequations} \label{EqFirstMomentMeasure}
	\begin{align} 
	  \!\!\!\!\!\!\!\!\!\!\!\!\mu^{(1)}_{\Phi}(B) &= \Exp\left[N_{\Phi}(B)\right] \label{EqFirstMomentMeasureA}
	  \\
	  &= \int \left(\sum_{x \in \varphi} 1_B(x)\right)P_{\Phi}(d\varphi) \label{EqFirstMomentMeasureB}
	  \\
	  &= \sum_{n \geqslant 0} \int \left(\sum_{1 \leqslant i \leqslant n} 1_B(x_i)\right)P_{\Phi}(dx_{1:n}), \label{EqFirstMomentMeasureC}
	\end{align}
      \end{subequations}
      where $x_{1:n} = \{x_1, \dots, x_n\}$, and
      \begin{subequations} \label{EqSecondMomentMeasure}
	\begin{align}
	  \mu^{(2)}_{\Phi}(B, B') &= \Exp\left[N_{\Phi}(B)N_{\Phi}(B')\right] \label{EqSecondMomentMeasureA}
	  \\
	  &= \int \left(\sum_{x_i,x_j \in \varphi} 1_B(x_i)1_{B'}(x_j)\right)P_{\Phi}(d\varphi) \label{EqSecondMomentMeasureB}
	  \\
	  &= \sum_{n \geqslant 0} \int \left(\sum_{1 \leqslant i,j \leqslant n} 1_B(x_i)1_{B'}(x_j)\right)P_{\Phi}(dx_{1:n}). \label{EqSecondMomentMeasureC}
	\end{align}
      \end{subequations}
      The first moment measure $\mu^{(1)}_{\Phi}(B)$ provides the expected number of targets or \textit{mean target number} inside $B$, while $\mu^{(2)}_{\Phi}(B, B')$ denotes the joint expectation of the target number inside $B$ and $B'$.
      
      Note that, $B$ and $B'$ can be selected such that they overlap\footnote{In this case, the realisations of $\Phi$ with targets in $B \cap B'$ will have non-zero values for both $N_{\Phi}(B)$ and $N_{\Phi}(B')$. Consequently, the inner summation in \eqref{EqSecondMomentMeasureC} will have non-zero terms for $i=j$.}, i.e., $B \cap B' \neq \emptyset$. In particular, the \textit{variance} $\var_{\Phi}$ of the point process $\Phi$ \cite{Stoyan_D_1995} in any region $B \in \Bbold_{\Xcal}$ is defined by
      \begin{equation} \label{EqVariance}
	\var_{\Phi}(B) = \mu^{(2)}_{\Phi}(B, B) - \left[\mu^{(1)}_{\Phi}(B)\right]^2.
      \end{equation}
      Note that the variance is a function, but not a \textit{measure}, on the Borel $\sigma$-algebra $\Bbold_{\Xcal}$. It does not necessarily admit a density, in general, even if $\mu^{(2)}_{\Phi}$ and $\mu^{(1)}_{\Phi}$ do. This fact motivates the measure-theoretical approach adopted throughout this article.
      
      The \textit{regional statistics} $(\mu^{(1)}_{\Phi}(B), \var_{\Phi}(B))$ provide an \textit{approximate} description of $N_{\Phi}(B)$, i.e. the \textit{local} number of target in B according to the point process $\Phi$:
      \begin{itemize}
	\item $\mu^{(1)}_{\Phi}(B)$ is the mean target number within $B$;
	\item $\var_{\Phi}(B)$ quantifies the dispersion of the target number within $B$ around its mean value.
      \end{itemize}
      \noindent
      Note that higher-order moments of a point process can be defined -- from the joint expectation of random variables $N_{\Phi}(B)$ as for the variance \eqref{EqSecondMomentMeasure} -- in order to provide a more complete description of the target number inside $B$. Derivation of such higher-order statistics is left out of the scope of this article.
      
    \subsection{Functional differentiation} \label{SubsecFunctionalDifferentiation}
      Statistical quantities describing a point process can be extracted through differentiation of various functionals, such as its \textit{probability generating functional} (PGFl) or its \textit{Laplace functional} (see Section \ref{SubsecGeneratingFunctionals}). Several functional differentials may be defined. Moyal used the G\^{a}teaux differential \cite{Hille_E_1957} in his early study on point processes \cite{Moyal_JE_1962}; although it is endowed with a sum and a product rule similar to ordinary differentials of real-valued functions, it lacks a chain (or composition) rule that would facilitate the derivation of multi-object filtering equations.
      
      In this article we exploit the multi-object filtering framework in \cite{Clark_DE_2012_2, Clark_DE_2013_2}, which considers the \textit{chain differential} \cite{Bernhard_P_2006}, in order to prove the results we present in Section \ref{SecFilter}. A restriction of the  G\^{a}teaux differential, the chain differential admits a composition rule. The chain differential $\delta F(h; \eta)$ of a functional $F$, (evaluated) at function $h$ in the direction (or increment) $\eta$, is defined as
      \begin{equation} \label{EqFunctionalDifferentiation}
	\delta F(h; \eta) = \lim_{n \rightarrow \infty} \frac{F(h + \epsilon_n\eta_n ) - F(h)}{\epsilon_n},
      \end{equation}
      where $\{\eta_n\}_{n \geqslant0}$ is a sequence of functions $\eta_n$ converging (pointwise) to $\eta$, $\{\epsilon_n\}_{n \geqslant0}$ is a sequence of positive real numbers converging to zero, if the limit exists and is identical for any admissible sequences $\{\eta_n\}_{n \geqslant0}$ and $\{\epsilon_n\}_{n \geqslant0}$ \cite{Bernhard_P_2006}. An example of chain differentiation for multi-object filtering is given in \cite{Clark_DE_2012}.
      
    \subsection{Generating functionals} \label{SubsecGeneratingFunctionals}
      The PGFl of a point process $\Phi$ is defined by the expectation
      \begin{subequations} \label{EqPGFlTargetProcess}
	\begin{align}
	  \Gcal_{\Phi}[h] &= \Exp\left[\prod_{x \in \Phi} h(x)\right] \label{EqPGFlTargetProcessA}
	  \\
	  &= \int \left(\prod_{x \in \varphi} h(x)\right)P_{\Phi}(d\varphi) \label{EqPGFlTargetProcessB}
	  \\
	  &= \sum_{n \geqslant 0} \int \left(\prod_{i = 1}^n h(x_i)\right)P_{\Phi}(dx_{1:n}), \label{EqPGFlTargetProcessC}
	\end{align}
      \end{subequations}
      where $h$ is a test function, i.e., a real-valued function belonging to the space of bounded measurable functions on $\Xcal$, such that $0 \leqslant h(x) \leqslant 1$ and $1 - h$ vanishes outside some bounded region of $\Xcal$ \cite{VereJones_D_2008}.
      
      The Laplace functional \cite{VereJones_D_2008, Stoyan_D_1995} of a point process $\Phi$ is given by the expectation
      \begin{subequations} \label{EqLaplaceFunctionalTargetProcess} 
	\begin{align}
	  \Lcal_{\Phi}[f] &= \Exp\left[\prod_{x \in \Phi} e^{-f(x)}\right] \label{EqLaplaceFunctionalTargetProcessA}
	  \\
	  &= \int \exp\left(-\sum_{x \in \varphi} f(x)\right)P_{\Phi}(d\varphi) \label{EqLaplaceFunctionalTargetProcessB}
	  \\
	  &=  \sum_{n \geqslant 0} \int \exp\left(-\sum_{i = 1}^n f(x_i)\right) P_{\Phi}(dx_{1:n}). \label{EqLaplaceFunctionalTargetProcessC}
	\end{align}
      \end{subequations}
      Both functionals fully characterise the probability distribution $P_{\Phi}$ and are linked by the relation
      \begin{equation} \label{EqRelationFunctionals}
	\Lcal_{\Phi}[f] = \Gcal_{\Phi}[e^{-f}].
      \end{equation}
      The probability distribution and the \textit{factorial} moment measures of a point process can easily be retrieved from functional differentials of the PGFl, making the PGFl a popular tool in multi-object filtering. Mahler's original construction of the PHD \cite{Mahler_RPS_2003} and CPHD \cite{Mahler_RPS_2007} filters, for example, exploits the differentiated PGFl. In our derivations for the second-order moment measure, we use 
      \textit{non-factorial} moment measures which are  easily retrieved from the Laplace functional \cite{Stoyan_D_1995}. To be precise, the \textit{factorial} moment measures $\alpha^{(n)}$ have a different construction and definition than the \textit{non-factorial} moment measures $\mu^{(n)}$ and will not be considered further in this article with the notable exception of the first factorial moment measure $\alpha^{(1)}$, which coincides with the first (non-factorial) moment measure $\mu^{(1)}$.
      
      The first and second moment measures of a point process $\Phi$ in any regions $B, B' \in \Bbold_{\Xcal}$ are given by the differentials \cite{Stoyan_D_1995}
      \begin{align} 
	\mu^{(1)}_{\Phi}(B) &= \left.\delta(\Gcal_{\Phi}[h]; 1_{B})\right|_{h = 1}, \label{EqFirstMomentComputation}
	\\
	\mu^{(2)}_{\Phi}(B, B') &= \left.\delta^2(\Lcal_{\Phi}[f]; 1_{B}, 1_{B'})\right|_{f = 0}, \label{EqSecondMomentComputation}
      \end{align}
      where $1_{B}$ is the indicator function on $B$
      \begin{equation} \label{EqIndicatorFunction}
	1_{B}(x) =
	\left\{
	  \begin{aligned}
	    &1 &\textnormal{if~}x \in B,
	    \\
	    &0 &\textnormal{if~}x \notin B.
	  \end{aligned}
	\right.
	, \quad\quad x \in \Xcal. 
      \end{equation}
      \noindent
      For the sake of simplicity, the superscript on the first moment measures is omitted in the rest of the article and $\mu^{(1)}_{\Phi}$ is denote by $\mu_{\Phi}$.
      
    \subsection{Multi-target Bayesian filtering} \label{SubsecBayesFiltering}
      In multi-object detection and tracking problems, the \textit{target process} $\Phi_{k|k}$ is a point process providing a stochastic description of the posterior distribution of the targets in the state space at time $k > 0$, based on the measurement history up to time $k$.
      
      Bayesian filtering principles are applicable to the multi-object framework \cite{Mahler_RPS_2007}. The law of the filtered state $P_{\Phi_{k|k}}$ is updated through sequences of \textit{prediction steps} -- according (acc.) to target birth, motion, and death models -- and \textit{data update steps} -- acc. to the current set of measurements\footnote{Each measurement has an individual state in the observation space $\Zcal \subset \Rset^{d_z}$ and $E_{\Zcal}$ is the space of all the sets of points in $\Zcal$.} $z_{1:m}^k \in E_{\Zcal}$. The full multi-target Bayes' filter reads as follows \cite{Mahler_RPS_2007_3}:
      \begin{align}
	P_{\Phi_{k|k-1}}(d\xi) &= \int T_{k|k-1}(d\xi|\varphi) P_{\Phi_{k-1|k-1}}(d\varphi), \label{EqMultiBayesFilter1}
	\\
	P_{\Phi_{k|k}}(d\xi|z_{1:m}^k) &= \frac{L_k(z_{1:m}^k|\xi)P_{\Phi_{k|k-1}}(d\xi)}{\int L_k(z_{1:m}^k|\varphi) P_{\Phi_{k|k-1}}(d\varphi)}, \label{EqMultiBayesFilter2}
      \end{align}
      where $T_{k|k-1}$ is the Markov transition kernel between time steps $k - 1$ and $k$, and $L_k$ is the multi-measurement/multi-target likelihood at time step $k$ (detailed later)\footnote{In the scope of this article, the infinitesimal neighbourhoods $dx_{1:n}$ defined around any point $x_{1:n} \in \Xcal^n$ are always chosen as elements of the product Borel $\sigma$-algebra $\Bbold_{\Xcal}^{\otimes n}$. Thus, $P(dx_{1:n}) = Q(dx_{1:n})$ is a notation for the well-defined expression $\int f(x_{1:n})P(dx_{1:n}) = \int f(x_{1:n})Q(dx_{1:n})$ for any test function $f$.}.
      
      Equivalent expression of the multi-target Bayes' filter can be provided through generating functionals. The PGFls of the predicted $\Phi_{k|k-1}$ and updated $\Phi_{k|k}$ processes are\cite{Clark_DE_2012_2}:
      \begin{align}
	\Gcal_{\Phi_{k|k-1}}[h] &= \iint \Bigg(\prod_{x \in \xi} h(x)\Bigg) T_{k|k-1}(d\xi|\varphi) P_{\Phi_{k-1|k-1}}(d\varphi), \label{EqMultiBayesPGFlFilter1}
	\\
	\Gcal_{\Phi_{k|k}}[h|z_{1:m}^k] &= \frac{\int \left(\prod_{x \in \varphi} h(x)\right) L_k(z_{1:m}^k|\varphi)P_{\Phi_{k|k-1}}(d\varphi)}{\int L_k(z_{1:m}^k|\varphi)P_{\Phi_{k|k-1}}(d\varphi)}. \label{EqMultiBayesPGFlFilter2}
      \end{align}
      Using \eqref{EqRelationFunctionals}, we can write an equivalent expression with the Laplace functionals:
      \begin{align}
	\Lcal_{\Phi_{k|k-1}}[f] &= \iint e^{-\sum_{x \in \xi} f(x)} T_{k|k-1}(d\xi|\varphi) P_{\Phi_{k-1|k-1}}(d\varphi), \label{EqMultiBayesLaplaceFilter1}
	\\
	\Lcal_{\Phi_{k|k}}[f|z_{1:m}^k] &= \frac{\int e^{-\sum_{x \in \xi} f(x)} L_k(z_{1:m}^k|\varphi)P_{\Phi_{k|k-1}}(d\varphi)}{\int L_k(z_{1:m}^k|\varphi)P_{\Phi_{k|k-1}}(d\varphi)}. \label{EqMultiBayesLaplaceFilter2}
      \end{align}
      \noindent
      For the sake of tractability, assumptions are often made on the prior $\Phi_{k-1|k-1}$ and/or the predicted $\Phi_{k|k-1}$ processes which subsequently lead to closed-form expressions of specific filters propagating \textit{incomplete information}.
    
  \section{The PHD and the CPHD filters with regional variance in target number} \label{SecFilter}
    In this section, we aim to provide the regional statistics of the updated target process for the CPHD and the PHD filters. We review both filters and identify the updated process from which we wish to produce the statistics in Section \ref{SubsecPrinciple}. We then provide the expression of its first (Section \ref{SubsecFirstMomentMeasure}) and second (Section \ref{SubsecSecondMomentMeasure}) moment measures for both filters. The main results of this article, the regional statistics for the CPHD and the PHD filters, follow in Section \ref{SubsecMainResult}. We discuss the procedures to extract the regional statistics for the Sequential Monte Carlo (SMC) implementations of the CPHD and PHD filters in Section \ref{SubsecImplementation}.
    
    The expressions of the first moment measures are well established results from the usual PHD \cite{Mahler_RPS_2003} and the CPHD \cite{Mahler_RPS_2007} filters. The derivation presented in this article, however, exploits the recent framework proposed in \cite{Clark_DE_2012_2}. On the other hand, the expression of the second moment measure is a novel result exposed in the authors' recent conference papers \cite{Delande_E_2013, Delande_E_2013_2}.
      
    \subsection{Principle} \label{SubsecPrinciple}
      The PHD \cite{Mahler_RPS_2003} and the CPHD \cite{Mahler_RPS_2007} filters are perhaps the most popular approximations to the multi-target Bayes' filter \eqref{EqMultiBayesFilter1}, \eqref{EqMultiBayesFilter2}. The predicted target process $\Phi_{k|k-1}$ is either approximated by an independent and identically distributed (i.i.d.) process (CPHD filter), or by a Poisson process (PHD filter).
      
\begin{figure}[b]
	\centering
	\includegraphics[width=0.80\columnwidth]{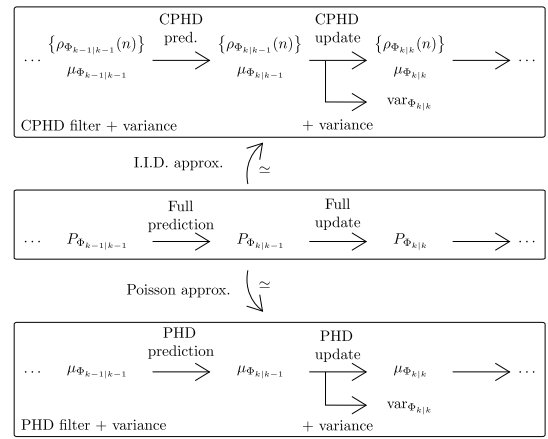}
	\caption{PHD and CPHD filtering with variance.\label{FigFilterVariance}}
      \end{figure}

      An i.i.d. process \cite{Vo_BT_2008_2} is completely described by 1) its cardinality distribution $\rho_{\Phi}$\footnote{$\rho_{\Phi}(n)$ is the probability that a realisation $\varphi$ of the point process $\Phi$ has size $n$, i.e. the probability that there are exactly $n$ targets in the surveillance scene.}, and 2) its first moment measure\footnote{An i.i.d. process $\Phi$ is usually described by the Radon-Nikodym derivative of its first moment measure $\mu_{\Phi}$ w.r.t. to the Lebesgue measure, also called its \textit{first moment density} $v_{\Phi}$ or \textit{intensity} or \textit{Probability Hypothesis Density} \cite{Mahler_RPS_2003}. Since we are interested in producing higher-order statistics on the target number, i.i.d. processes on targets are described by their first moment measure $\mu_{\Phi}$ instead. I.i.d processes on measurements, however, are still described by their intensity $v_{\Phi}$ or, to be precise, by their normalised intensity or spatial distribution (see Theorem \ref{TheoCPHDStatistics} and \ref{TheoPHDStatistics}).} $\mu_{\Phi}$. Hence, the CPHD filter propagates a cardinality distribution $\rho_{\Phi}$ and a moment measure $\mu_{\Phi}$. A Poisson process is a specific case of an i.i.d. process in which the cardinality distribution is a Poisson distribution with rate $\mu_{\Phi}(\Xcal) = \int \mu_{\Phi}(dx)$. Hence, a Poisson process is completely described by its first moment measure $\mu_{\Phi}$, propagated by the PHD filter (see Figure \ref{FigFilterVariance}).
      
      The updated target process $\Phi_{k|k}$ is \textit{not}, in the general case, i.i.d. (respectively Poisson) even if the predicted $\Phi_{k|k-1}$ is; that is, the updated probability distribution $P_{\Phi_{k|k}}$ is \textit{not} completely described by the output of the CPHD (respectively PHD) filter. As a consequence, the computation of the variance $\var_{\Phi_{k|k}}$ provides \textit{additional} information on the updated process $\Phi_{k|k}$, before its collapse into a i.i.d. (respectively Poisson) process in the next time step (see Figure \ref{FigFilterVariance}).
      
      As shown in Figure \ref{FigFilterVariance}, this article focuses on the generation of additional information describing the \textit{updated} target process; hence, the prediction step \eqref{EqMultiBayesPGFlFilter1} will not be further mentioned. The rest of the article describes the extraction of the information statistics $(\mu_{\Phi_{k|k}}, \var_{\Phi_{k|k}})$ at an arbitrary time step $k > 0$. For the sake of simplicity, we discard the time subscripts and denote the predicted and the update processes with $\Phi$ and $\Phi_+$ respectively. In addition, we denote the current set of measurements by $z_{1:m}$.

    \subsection{First moment measure (CPHD and PHD updates)} \label{SubsecFirstMomentMeasure}
      \begin{lemma}\textnormal{First moment measure (CPHD update) \cite{Mahler_RPS_2007}, \cite{Vo_BT_2007}} \label{LemCPHDFirstMoment} \newline
	The first moment measure of the updated process $\Phi_{+}$ in any region $B \in \Bbold_{\Xcal}$, under the assumptions that \cite{Mahler_RPS_2007}:
	\begin{enumerate}
	  \item The predicted process $\Phi$ is an i.i.d. process, with cardinality distribution $\rho_{\Phi}$ and first moment measure $\mu_{\Phi}$;
	  \item A target $x$ is detected by the sensor with probability $p_d(x)$;
	  \item If detected, a target $x$ produces a single measurement $z$ acc. to the single-measurement/single-target likelihood $\hat{L}(z|x)$;
	  \item The clutter is an i.i.d. process, with cardinality distribution $\rho_c$ and spatial distribution $c(\cdot)$;
	\end{enumerate}
	is given by
	\begin{equation} \label{EqCPHDFirstMoment}
	  \mu_{\Phi_{+}}(B) = \mu^{\phi}_{\Phi}(B)\ell_1(\phi) + \sum_{\mathclap{z \in z_{1:m}}}~ \frac{\mu^z_{\Phi}(B)}{c(z)}\ell_1(z),
	\end{equation}
	where the corrector terms $\ell_1(\phi)$ and $\ell_1(z)$ are given by
	\begin{align}
	  &\left\{
	    \begin{aligned}
	      \ell_1(\phi) &= \frac{\left<\Upsilon^1[\mu_{\Phi},z_{1:m}], \rho_{\Phi}\right>}{\left<\Upsilon^0[\mu_{\Phi},z_{1:m}], \rho_{\Phi}\right>},
	      \\
	      \ell_1(z) &= \frac{\left<\Upsilon^1[\mu_{\Phi},z_{1:m} \setminus z], \rho_{\Phi}\right>}{\left<\Upsilon^0[\mu_{\Phi},z_{1:m}], \rho_{\Phi}\right>},
	    \end{aligned}
	  \right. \label{EqCPHDFirstMomentCorrector}
	\end{align}
	where (following the notation introduced by Vo, et. al. in \cite{Vo_BT_2007}):
	\begin{align}
	  &\Upsilon^u[\mu_{\Phi},Z](n) \nonumber
	  \\
	  &= \sum_{d = 0}^{\mathclap{\min(|Z|, n)}}~ \frac{n!(|Z| - d)!}{(n - (d + u))!} \rho_c(|Z| - d) \frac{\mu^{\phi}_{\Phi}(\Xcal)^{n - (d + u)}}{\mu_{\Phi}(\Xcal)^n} e_d(Z), \label{EqVoNotation1}
	  \\
	  &\left<\Upsilon^u[\mu_{\Phi},Z], \rho_{\Phi}\right> = \sum_{n \geqslant 0} \Upsilon^u[\mu_{\Phi},Z](n)\rho_{\Phi}(n), \label{EqVoNotation2}
	\end{align}
	where for any region $B \in \Bbold_{\Xcal}$:
	\begin{align}
	  \mu^z_{\Phi}(B) &= \int 1_B(x)P(z|x)\mu_{\Phi}(dx), \label{EqKernelIntegrated1}
	  \\
	  \mu^{\phi}_{\Phi}(B) &= \int 1_B(x)P(\phi|x)\mu_{\Phi}(dx), \label{EqKernelIntegrated2}
	\end{align}
	where $P$ is the single-measurement/single-target observation kernel, i.e.
	\begin{align}
	  P(z|x) &= p_d(x)\hat{L}(z|x), \label{EqKernel1}
	  \\
	  P(\phi|x) &= 1 - p_d(x). \label{EqKernel2}
	\end{align}
	The function $e_d$ is the elementary symmetric function of order $d$ \cite{Vo_BT_2008_2}
	\begin{equation} \label{EqElementarySymmetricFunction}
	  e_d(\Xi) = \ds{\sum_{S \subseteq \Xi, |\Xi| = d}} \left(\ds{\prod_{\xi \in S}} \xi \right),
	\end{equation}
	applied in \eqref{EqVoNotation1} to the set $\left\{\frac{\mu^z_{\Phi}(\Xcal)}{c(z)}|z \in z_{1:m}\right\}$ and abusively noted $e_d(z_{1:m})$. 	
      \end{lemma}
      \noindent
      The proof is given in Appendix \ref{AppProofs} (Section \ref{AppLemCPHDFirstMoment}).

      \begin{corollary}\textnormal{First moment measure (PHD update) \cite{Mahler_RPS_2003}} \label{CorPHDFirstMoment} \newline
	The first moment measure of the updated process $\Phi_{+}$ in any region $B \in \Bbold_{\Xcal}$, under the assumptions given in Lemma \ref{LemCPHDFirstMoment} and the additional assumptions that \cite{Mahler_RPS_2003}:
	\begin{enumerate}
	  \item The predicted process $\Phi$ is Poisson;
	  \item The clutter is Poisson, whose rate is denoted by $\lambda_c$;
	\end{enumerate}
	is given by
	\begin{equation} \label{EqPHDFirstMoment}
	  \mu_{\Phi_{+}}(B) = \mu^{\phi}_{\Phi}(B) + \sum_{\mathclap{z \in z_{1:m}}}~ \frac{\mu^z_{\Phi}(B)}{\mu^{z}_{\Phi}(\Xcal) + \lambda_cc(z)}.
	\end{equation}
      \end{corollary}
      \noindent
      The proof is given in Appendix \ref{AppProofs} (Section \ref{AppCorPHDFirstMoment}).
      
    \subsection{Second moment measure (CPHD and PHD updates)} \label{SubsecSecondMomentMeasure}
      \begin{lemma}\textnormal{Second moment measure (CPHD update)} \label{LemCPHDSecondMoment} \newline
	Under the assumptions given in Lemma \ref{LemCPHDFirstMoment}, the second moment measure of the updated process $\Phi_{+}$ in any regions $B,~B' \in \Bbold_{\Xcal}$ is given by
	\begin{align}
	  &\mu^{(2)}_{\Phi_{+}}(B, B') \nonumber
	  \\
	  &= \mu_{\Phi_{+}}(B \cap B') + \mu^{\phi}_{\Phi}(B)\mu^{\phi}_{\Phi}(B')\ell_2(\phi) \nonumber
	  \\
	  &+ \mu^{\phi}_{\Phi}(B)\sum_{z \in z_{1:m}} \frac{\mu^{z}_{\Phi}(B')}{c(z)}\ell_2(z) + \mu^{\phi}_{\Phi}(B')\sum_{z \in z_{1:m}} \frac{\mu^{z}_{\Phi}(B)}{c(z)}\ell_2(z) \nonumber
	  \\
	  &+ \ds{\sideset{}{^{\neq}}\sum_{z, z' \in z_{1:m}}}\frac{\mu^{z}_{\Phi}(B)}{c(z)}\frac{\mu^{z'}_{\Phi}(B')}{c(z')}\ell_2(z, z'), \label{EqCPHDSecondMoment}
	\end{align}
	where the corrector terms $\ell_2(\phi)$, $\ell_2(z)$, and $\ell_2(z,z')$ are given by:
	\begin{align}
	  &\left\{
	    \begin{aligned}
	      \ell_2(\phi) &= \frac{\left<\Upsilon^2[\mu_{\Phi},z_{1:m}], \rho_{\Phi}\right>}{\left<\Upsilon^0[\mu_{\Phi},z_{1:m}], \rho_{\Phi}\right>}, 
	      \\
	      \ell_2(z) &= \frac{\left<\Upsilon^2[\mu_{\Phi},z_{1:m} \setminus z], \rho_{\Phi}\right>}{\left<\Upsilon^0[\mu_{\Phi},z_{1:m}], \rho_{\Phi}\right>},
	      \\
	      \ell_2(z, z') &= \frac{\left<\Upsilon^2[\mu_{\Phi},z_{1:m} \setminus \{z, z'\}], \rho_{\Phi}\right>}{\left<\Upsilon^0[\mu_{\Phi},z_{1:m}], \rho_{\Phi}\right>}.
	    \end{aligned}
	  \right. \label{EqCPHDSecondMomentCorrector}
	\end{align}
      \end{lemma}
      \noindent
      The proof is given in Appendix \ref{AppProofs} (Section \ref{AppLemCPHDSecondMoment}).
	
      \begin{corollary}\textnormal{Second moment measure (PHD update)} \label{CorPHDSecondMoment} \newline
	Under the assumptions given in Corollary \ref{CorPHDFirstMoment}, the second moment measure of the updated process $\Phi_{+}$ in any regions $B,~B' \in \Bbold_{\Xcal}$ is given by
	\begin{align}
	  &\mu^{(2)}_{\Phi_{+}}(B, B')\nonumber
	  \\
	  &= \mu_{\Phi_{+}}(B \cap B') + \mu^{\phi}_{\Phi}(B)\mu^{\phi}_{\Phi}(B') \nonumber
	  \\
	  &+ \mu^{\phi}_{\Phi}(B) \sum_{z \in z_{1:m}} \frac{\mu^{z}_{\Phi}(B')}{\mu^z_{\Phi}(\Xcal) + \lambda_cc(z)} \nonumber
	  \\
	  &+ \mu^{\phi}_{\Phi}(B') \sum_{z \in z_{1:m}} \frac{\mu^{z}_{\Phi}(B)}{\mu^z_{\Phi}(\Xcal) + \lambda_cc(z)} \nonumber
	  \\
	  &+ \ds{\sideset{}{^{\neq}}\sum_{z, z' \in z_{1:m}}}\frac{\mu^{z}_{\Phi}(B)}{\mu^z_{\Phi}(\Xcal) + \lambda_cc(z)}\frac{\mu^{z'}_{\Phi}(B')}{\mu^{z'}_{\Phi}(\Xcal) + \lambda_cc(z')}. \label{EqPHDSecondMoment}
	\end{align}
      \end{corollary}
      \noindent
      The proof is given in Appendix \ref{AppProofs} (Section \ref{AppCorPHDSecondMoment}).
      
    \subsection{Main results} \label{SubsecMainResult}
      The two following theorems are the main results of this article. Their proof is given in Appendix \ref{AppProofs} (Section \ref{AppTheoCPHDPHDStatistics}).
      
      \begin{theorem}\textnormal{Regional statistics (CPHD update)} \label{TheoCPHDStatistics} \newline
	Under the assumptions given in Lemma \ref{LemCPHDFirstMoment}, the regional statistics\footnote{Note (see Figure \ref{FigFilterVariance}) that the usual CPHD filter produces the updated cardinality distribution $\rho_{\Phi_{+}}$. Hence, it provides a full stochastic description of the target number \textit{in the whole state space}; that is, of the random variable $N_{\Phi_{+}}(\Xcal)$ (see Figure \ref{FigPointProcessAndCountingMeasure} with $B = \Xcal$). The regional variance can thus be extracted from the usual CPHD, but only for the specific region $B = \Xcal$.} of the updated process $\Phi_{+}$ in any region $B \in \Bbold_{\Xcal}$ are given by 
	\begin{align}
	  \mu_{\Phi_{+}}(B) &= \mu^{\phi}_{\Phi}(B) \ell_1(\phi) + \sum_{z \in z_{1:m}}\frac{\mu^z_{\Phi}(B)}{c(z)} \ell_1(z), \label{EqCPHDMeanResult}
	  \\
	  \var_{\Phi_{+}}(B) &= \mu_{\Phi_{+}}(B) + \mu^{\phi}_{\Phi}(B)^2\left[\ell_2(\phi) - \ell_1(\phi)^2\right] \nonumber
	  \\
	  &+ 2\mu^{\phi}_{\Phi}(B) ~\sum_{\mathclap{z \in z_{1:m}}}~ \frac{\mu^z_{\Phi}(B)}{c(z)} \left[\ell_2(z) - \ell_1(z)\ell_1(\phi)\right] \nonumber
	  \\
	  &+ \ds{\sum_{\mathclap{z, z' \in z_{1:m}}}}~ \frac{\mu^z_{\Phi}(B)}{c(z)} \frac{\mu^{z'}_{\Phi}(B)}{c(z')} \left[\ell_2^{\neq}(z,z') - \ell_1(z)\ell_1(z')\right], \label{EqCPHDVarianceResult}
	\end{align}
	where $\ell_2^{\neq}(z, z') = \ell_2(z, z')$ if $z \neq z'$, or zero otherwise.
      \end{theorem}
      
      \begin{theorem}\textnormal{Regional statistics (PHD update)\label{TheoPHDStatistics}} \newline
	Under the assumptions given in Corollary \ref{CorPHDFirstMoment}, the regional statistics of the updated process $\Phi_{+}$ in any region $B \in \Bbold_{\Xcal}$ are given by 
	\begin{align}
	  \mu_{\Phi_{+}}(B) &= \mu^{\phi}_{\Phi}(B) + \sum_{z \in z_{1:m}} \frac{\mu^z_{\Phi}(B)}{\mu^z_{\Phi}(\Xcal) + \lambda_cc(z)}, \label{EqPHDMeanResult}
	  \\
	  \var_{\Phi_{+}}(B) &= \mu^{\phi}_{\Phi}(B) \nonumber
	  \\
	  &+ \sum_{z \in z_{1:m}} \frac{\mu^z_{\Phi}(B)}{\mu^z_{\Phi}(\Xcal) + \lambda_cc(z)}\left(1 - \frac{\mu^z_{\Phi}(B)}{\mu^z_{\Phi}(\Xcal) + \lambda_cc(z)}\right). \label{EqPHDVarianceResult}
	\end{align}
      \end{theorem}
      
    \subsection{Discussion on implementation} \label{SubsecImplementation}
      We consider SMC implementations of the PHD and the CPHD filters and equip them with regional statistiscs. The resulting algorithms are given in Appendix~\ref{AppAlgorithms}.
      
      The SMC-PHD filter with regional variance can be easily drawn from the usual SMC-PHD filter \cite{Vo_BN_2005}. Indeed, the regional variance is computed using the terms that are already computed to find the regional mean \eqref{EqPHDMeanResult} in the SMC-PHD filter (see Algorithm~\ref{AlgPHD}). The computational complexity of the PHD filter with the variance is still linear w.r.t. the number of current measurements $m$.
      
      Similarly, the construction of the SMC-CPHD filter with regional variance is an extension to the well-known SMC-CPHD filter~\cite{Vo_BT_2008_2}. As shown in Algorithm~\ref{AlgCPHD}, the additional corrector terms $\ell_2(\phi)$, $\ell_2(z)$, and $\ell_2(z,z')$ \eqref{EqCPHDSecondMomentCorrector} are computed in parallel to the usual corrector terms $\ell_1(\phi)$ and $\ell_1(z)$ \eqref{EqCPHDFirstMomentCorrector}. In the usual CPHD filter, the bulk of the computational cost stems from the computation of $\ell_1(\phi)$ and $\ell_1(z)$ in the filtering equation \eqref{EqCPHDMeanResult} or, more specifically, the elementary symmetric functions \eqref{EqElementarySymmetricFunction} appearing in the $\Upsilon^0$ and $\Upsilon^1$ terms \eqref{EqVoNotation1}. The number of operations to compute $e_d(z_{1:m})$ is evaluated at $m\log^2m$ in \cite{Vo_BT_2007} and $m + 1$ elementary symmetric functions must be computed for $\ell_1(\phi)$ and $\ell_1(z)$. Thus, it has been shown by  Vo et al. that the computational complexity of the CPHD filter is $\Ocal(m^2\log^2 m)$, where $m$ is the number of current measurements \cite{Vo_BT_2007}.
      
      The corrector terms $\ell_2(\phi)$ and $\ell_2(z)$ \eqref{EqCPHDSecondMomentCorrector}, required for the computation of the regional variance \eqref{EqCPHDVarianceResult}, do not involve new elementary symmetric functions and can be found in parallel to $\ell_1(\phi)$ and $\ell_1(z)$ without significant additional cost (see Algorithm \ref{AlgCPHD}). On the other hand, $\ell_2(z,z')$ involves $\frac{m(m - 1)}{2}$ different $\Upsilon^2$ terms \eqref{EqVoNotation1} with additional elementary symmetric functions $e_d(z,z')$ -- for every couple of distinct measurements $z, z'$. Thus, the computational complexity of the SMC-CPHD filter with regional variance is $\Ocal(m^3\log^2 m)$.
    
  \section{Simulation examples} \label{SecSimulation}
\begin{figure}[H]
      \centering
      \includegraphics[width=0.7\columnwidth]{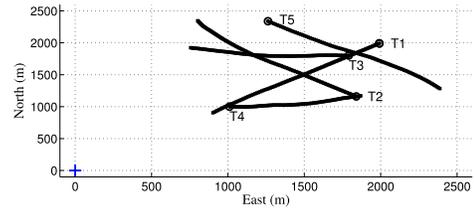}
      \caption[Example scenario: target trajectories (position plane) and sensor location]{Example scenario: target trajectories (position plane) and sensor location (`$\textcolor{blue}{+}$'). Circles indicate target initial positions.\label{FigScenarioTrajectories}}
    \end{figure}

    In this section, we demonstrate the concept of regional variance for the PHD and the CPHD filters using the multi-target scenario illustrated in \figurename~\ref{FigScenarioTrajectories}. A range-bearing sensor located at the origin takes measurements from five targets that appear and disappear over time in the surveillance scene. The sensor Field of View (FoV) is the circular region centred at the origin and with radius $\SI{3500}{\meter}$. The standard deviations in range and bearing are selected as $\SI{5}{\meter}$ and $\SI{1}{\degree}$ respectively. The clutter is generated from a Poisson process with rate $\lambda = 20$ and uniform over the FoV.
    
    The state of targets is described by a location $[x,y]$ and a velocity $[\dot{x},\dot{y}]$ component, and the subset of $\Rset^4$ that falls in the FoV is the state space $\Xcal$. The state transitions follow a linear constant velocity motion model and (slight) additive zero mean process noise after getting initiated with the values given in Table~\ref{TabTargets}. Trajectories of targets $1$ and $2$ cross each other at time $t = \SI{55}{\second}$.
    
    \begin{table}[H]
      \caption{Initial target states and track information\label{TabTargets}}
      \vspace{-4mm}
      \begin{center}
	\begin{tabular}{|c|c|c|}
	  \hline
	  Init. loc. ($\si{\meter}$) &  Init. vel. ($\si{\meter\per\second}$) & Time of birth/death ($\si{\second}$)\\
	  \hline
	  \hline
	  $[2000.0,\,2000.0]^T$ & $[-9.1,\, -9.1]^T$ & $0/110$
	  \\
	  $[1850.0,\,4000.0]^T$ & $[-10.0,\, -10.0]^T$ & $20/130$
	  \\
	  $[1800.0,\,1800.0]^T$ & $[-10.0,\, 0.0]^T$ & $40/150$
	  \\
	  $[1000.0,\,1000.0]^T$ & $[10.0,\, 0.0]^T$ & $70/170$
	  \\
	  $[1250.0,\,2350.0]^T$ & $[12.0,\, -12.0]^T$ & $90/190$
	  \\
	  \hline
	\end{tabular}\newline
      \end{center}
    \end{table}
    
    \subsection{Variance as a global statistic} \label{SubsecGlobalVariance}
      In this example, we consider the regional variance over the FoV under different target detection probabilities. Doing so, we demonstrate the effect of the probability of detection $p_d$ on the uncertainty of the estimated target number. We simulate measurements with $p_d = 0.95,0.90$, and $0.85$ and run both the CPHD and the PHD filters. The mean and the variance in the target number within the FoV (given by the regional statistics evaluated in the whole FoV) are computed using Algorithms~\ref{AlgPHD}~and~\ref{AlgCPHD}.
      
\begin{figure}
	\centering
	\subfloat[]
	{
	  \includegraphics[width=0.32\columnwidth]{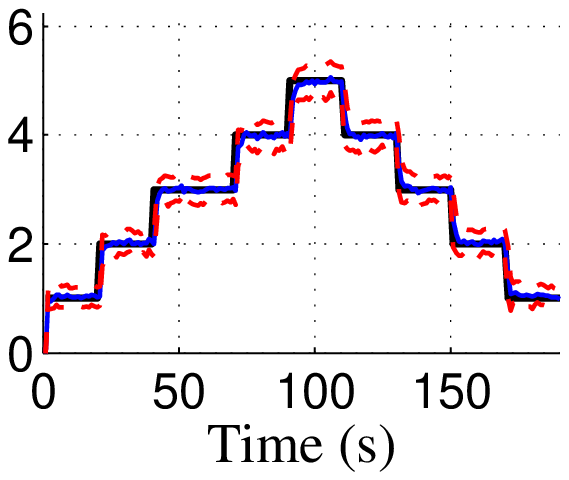}
	  \label{SubFigCPHDFOVpd095}
	}
	\subfloat[]
	{
	  \includegraphics[width=0.32\columnwidth]{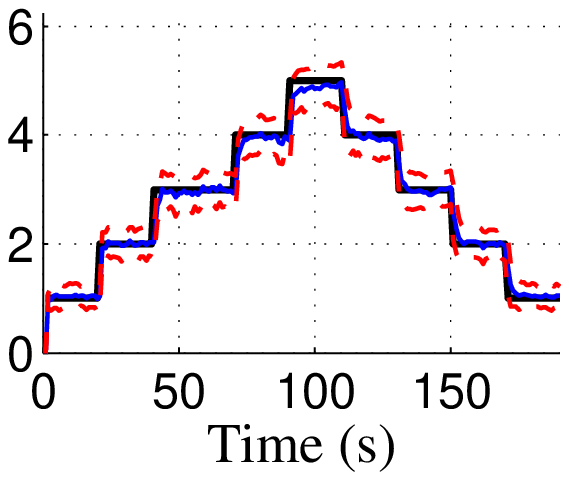}
	  \label{SubFigCPHDFOVpd090}
	}
	\subfloat[]
	{
	  \includegraphics[width=0.32\columnwidth]{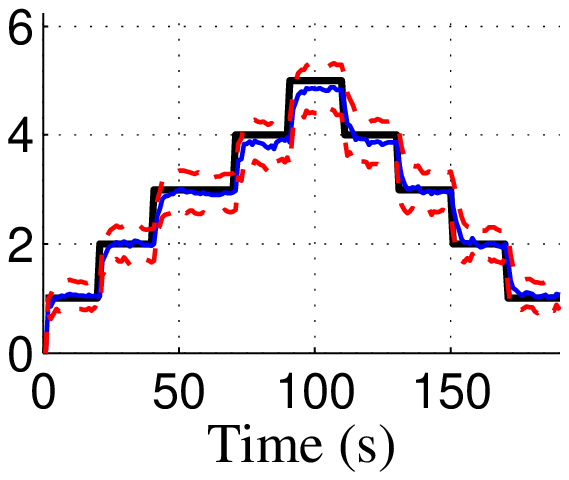}
	  \label{SubFigCPHDFOVpd085}
	}
	\\
	\subfloat[]
	{
	  \includegraphics[width=0.32\columnwidth]{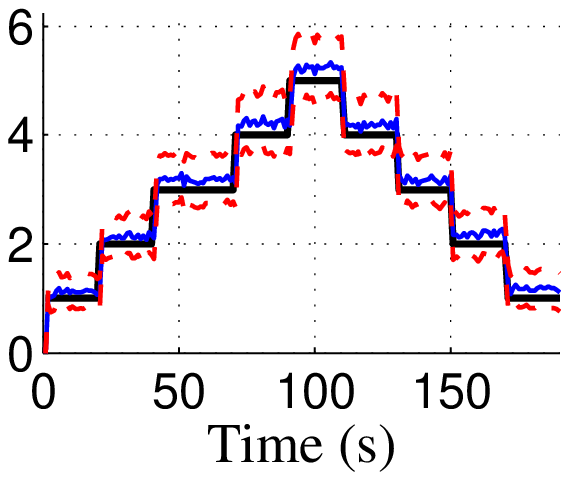}
	  \label{SubFigPHDFOVpd095}
	}
	\subfloat[]
	{
	  \includegraphics[width=0.32\columnwidth]{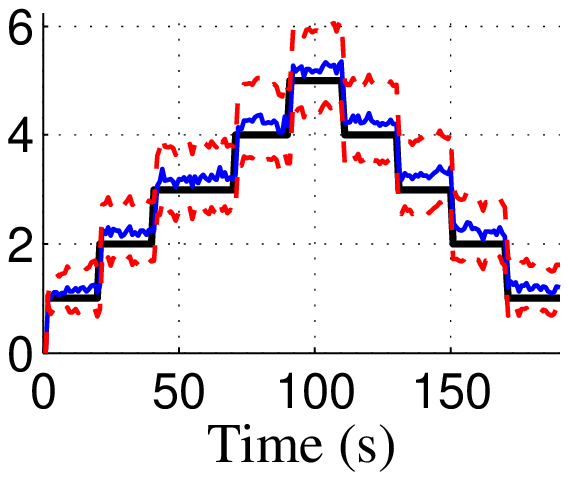}
	  \label{SubFigPHDFOVpd090}      
	}
	\subfloat[]
	{
	  \includegraphics[width=0.32\columnwidth]{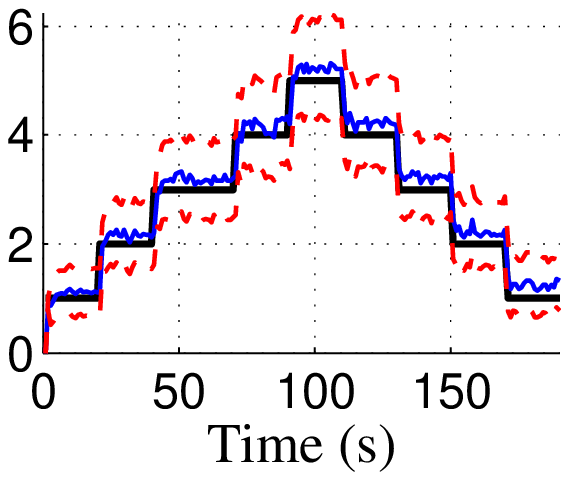}
	  \label{SubFigPHDFOVpd085}
	}
	\caption[Regional variance in the FoV with the CPHD and the PHD filters]{Mean target number and $\pm 1$ standard deviation (square root of the regional variance) integrated in the whole FoV, for $p_d = 0.95$, $0.90$, and $0.85$. Results obtained using  \subref{SubFigCPHDFOVpd095}--\subref{SubFigCPHDFOVpd085} the CPHD filter, and, \subref{SubFigPHDFOVpd095}--\subref{SubFigPHDFOVpd085} the PHD filter. The plots are the averages over 100 Monte Carlo runs.\label{FigCPHDFOV}}
      \end{figure}

      In \figurename~\ref{FigCPHDFOV}\subref{SubFigCPHDFOVpd095}--\subref{SubFigCPHDFOVpd085}, we present the mean target number in the FoV (blue line) computed using the CPHD filter, together with the ground truth (black line). The variance in target number within the FoV is used to quantify the level of uncertainty in the mean target number. Specifically, we present confidence intervals as the $\pm 1$ square root of the regional variance which in turn admits a standard deviation interpretation. We note that the uncertainty increases as we lower the probability of detection, coinciding with our intuition. The behaviour of the confidence bounds computed using the PHD filter is similar as seen in \figurename~\ref{FigCPHDFOV}\subref{SubFigPHDFOVpd095}--\subref{SubFigPHDFOVpd085}.
      
\begin{figure}
	\centering
	\subfloat[]
	{
	  \includegraphics[width=0.7\columnwidth]{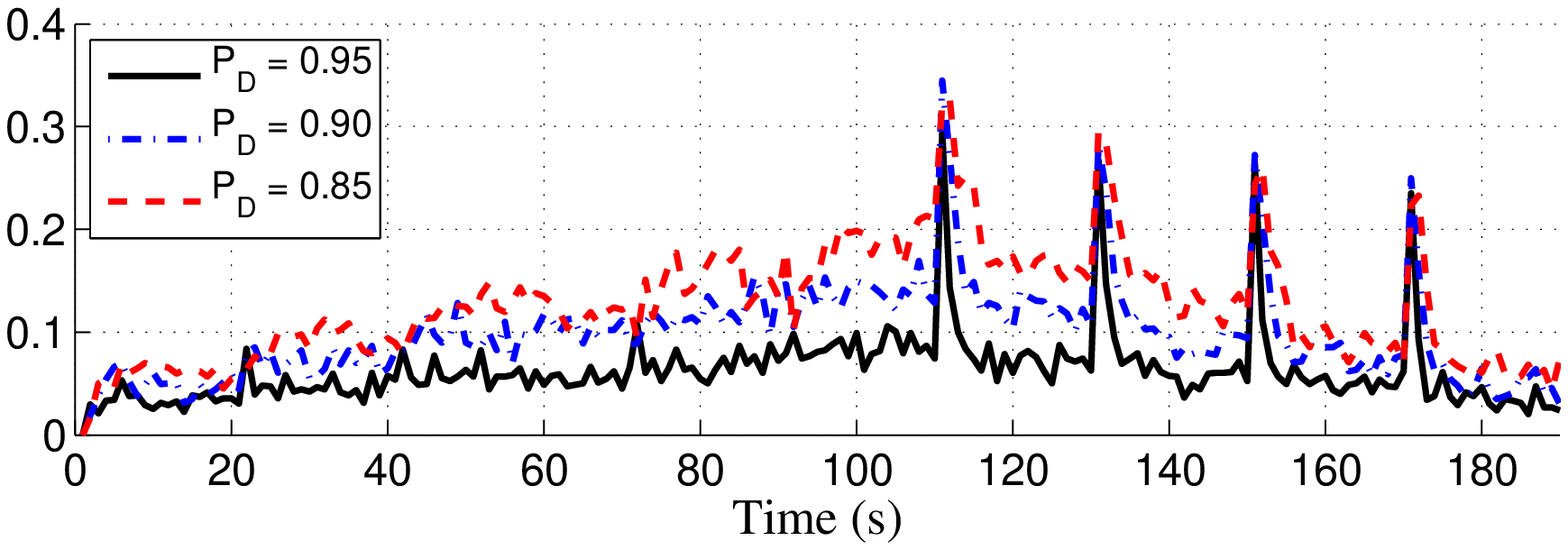}
	  \label{SubFigCPHDFOVAll}
	}
	\\
	[-2pt]
	\subfloat[]
	{
	  \includegraphics[width=0.7\columnwidth]{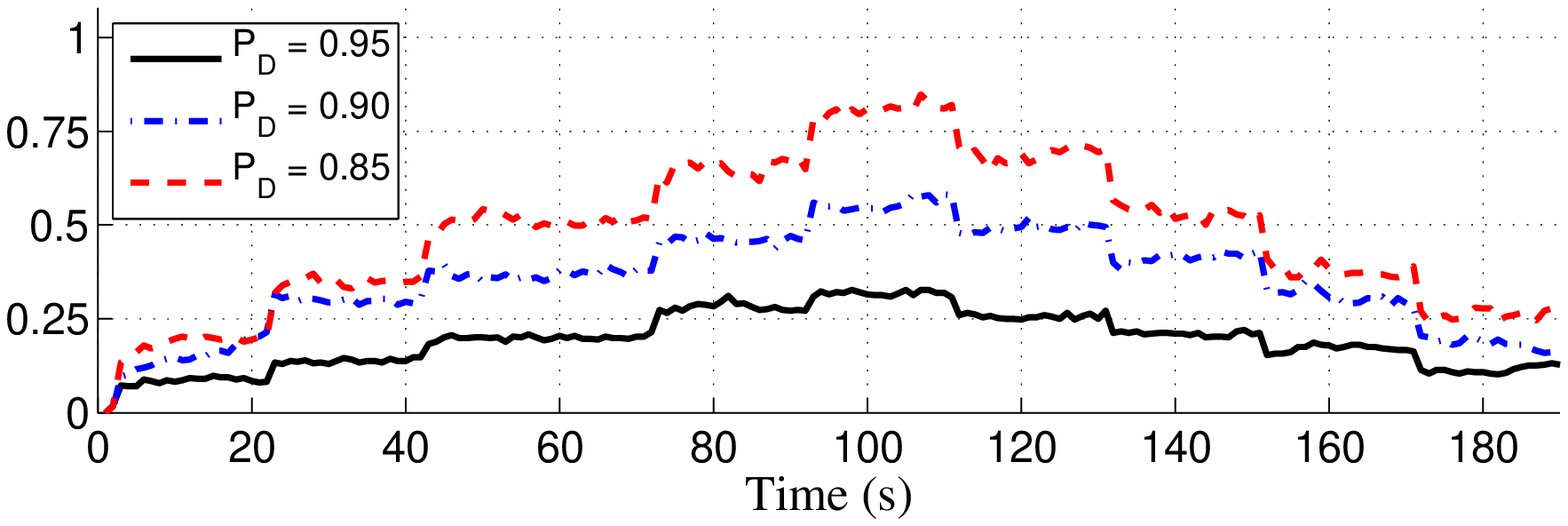}
	  \label{SubFigPHDFOVAll}
	}
	\caption[Regional variance in the FoV with the CPHD and PHD filters]{Regional variance, integrated in the whole FoV \subref{SubFigCPHDFOVAll} using the CPHD, \subref{SubFigPHDFOVAll} the PHD  filter, for $p_d = 0.95$, $0.90$, and $0.85$. The plots are the averages over 100 Monte Carlo runs.\label{FigFOVAll}}
      \end{figure}

      The regional variances used to find the aforementionned confidence intervals are presented in \figurename~\ref{FigFOVAll}. In \figurename~\ref{FigFOVAll}\subref{SubFigCPHDFOVAll}, we plot the results obtained using the CPHD filter as $p_d$ goes from $0.95$ to $0.85$. Similar plots for the PHD filter are provided in \figurename~\ref{FigFOVAll}\subref{SubFigPHDFOVAll}. The increasing uncertainty with the decreasing $p_d$ can clearly be seen. We also note that the variance over the FoV grows significantly more with the PHD than with the CPHD filter as $p_d$ is lowered.
      
    \subsection{Variance as a local statistic} \label{SubsecLocalVariance}
\begin{figure}
	\centering
	\includegraphics[width=0.7\columnwidth]{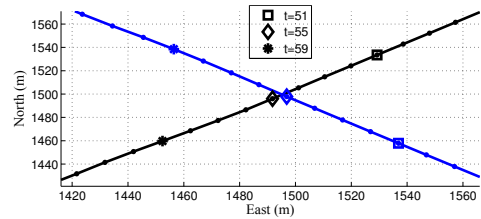}
	\caption[Approaching targets]{Approaching targets: targets $1$ (black) and $2$ (blue) crisscrossing around time step $t = \SI{55}{\second}$. The distance between the targets is $76.1, 5.4$ and $78.9$ at time steps $t = \SI{51}{\second}$, $\SI{55}{\second}$, and $\SI{59}{\second}$, respectively.\label{FigApproachingTargets}}
      \end{figure}

      In this example, we illustrate the variance evaluated in regions of various sizes within the FoV. Specifically, we consider concentric circular regions of growing radius around the location of target $1$ while its trajectory crosses that of target~$2$~(\figurename~\ref{FigApproachingTargets}). We vary the radiuses from $r=\SI{1}{\meter}$ to $\SI{200}{\meter}$ with $\SI{1}{\meter}$ steps at time steps $t=51,55$ and $\SI{59}{\second}$. The distance between the targets are $76.1, 5.4$ and $\SI{78.9}{\meter}$., respectively, at these time instants, so, the regions with larger radius cover both targets.
      
      We compute both the mean target number in these concentric regions and the associated uncertainty quantified by the proposed regional variance. We expect the mean target number to be monotonically increasing as a function of the radius and to reach approximately two for the larger circles. The regional variance, on the other hand, is not necessarily monotonic and we expect its envelope to be an indicator of whether target $1$ can be resolved in the sense that we can identify circular regions that contain {\it only} target $1$ with high confidence.
      
\begin{figure}
	\centering
	\subfloat[]
	{
	  \includegraphics[width=0.32\columnwidth]{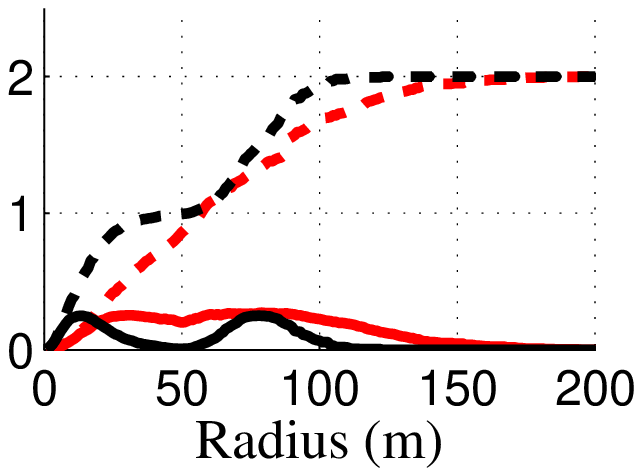}
	  \label{SubFigCPHDt51}
	}
	\subfloat[]
	{
	  \includegraphics[width=0.32\columnwidth]{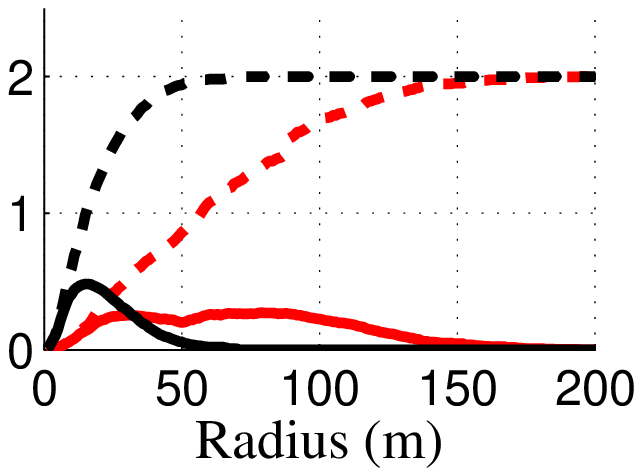}
	  \label{SubFigCPHDt55}
	}
	\subfloat[]
	{
	  \includegraphics[width=0.32\columnwidth]{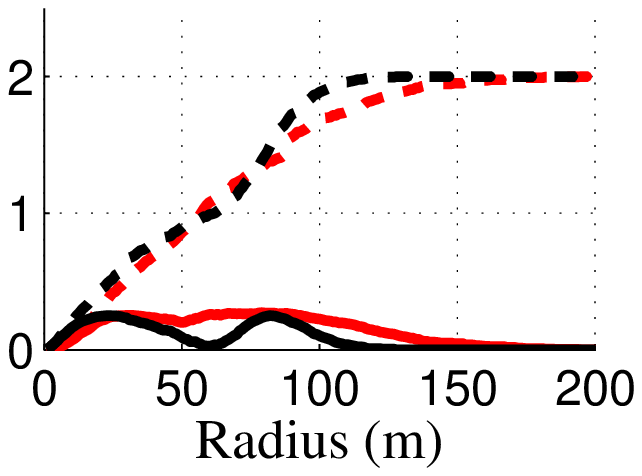}
	  \label{SubFigCPHDt59}
	}\\
	\subfloat[]
	{
	  \includegraphics[width=0.32\columnwidth]{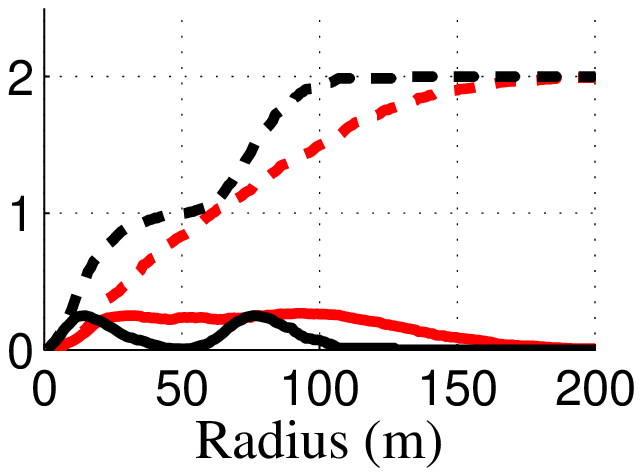}
	  \label{SubFigPHDt51}
	}
	\subfloat[]
	{
	  \includegraphics[width=0.32\columnwidth]{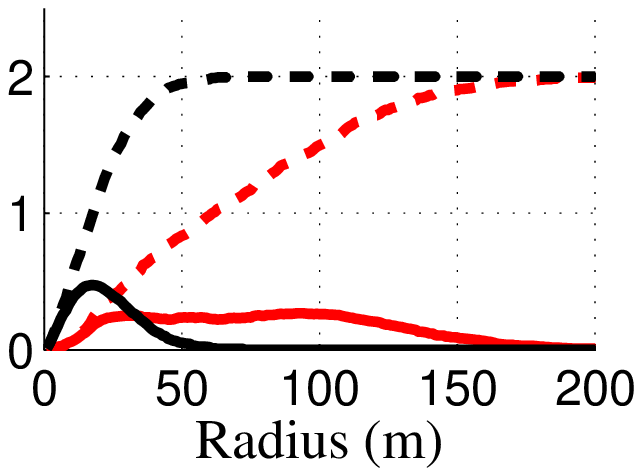}
	  \label{SubFigPHDt55}
	}
	\subfloat[]
	{
	  \includegraphics[width=0.32\columnwidth]{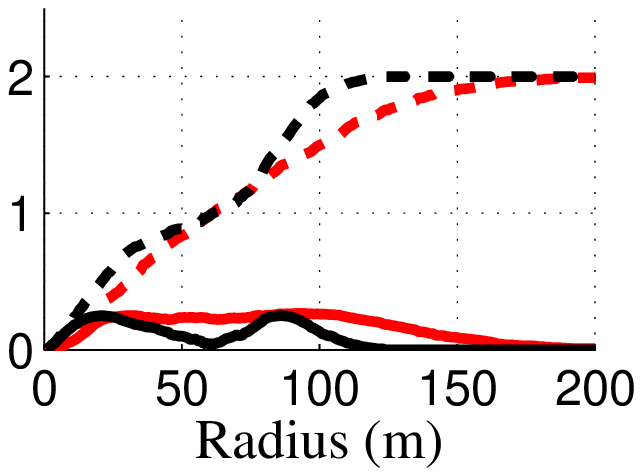}
	  \label{SubFigPHDt59}
	}
	\caption[Regional mean and variance in circular regions around target one.]{Regional mean (plain lines) and variance (dotted lines) in circular regions centred at the position of target $1$ at time $t=51, 55$ and $\SI{59}{\second}$ for the CPHD \subref{SubFigCPHDt51}--\subref{SubFigCPHDt59} and the PHD \subref{SubFigPHDt51}--\subref{SubFigPHDt59} filters, respectively. Results are given for a superior (black lines) and an inferior (red lines) range-bearing sensor.\label{FigCPHDApproaching}}
      \end{figure}

      In \figurename~\ref{FigCPHDApproaching}\subref{SubFigCPHDt51}--\subref{SubFigCPHDt59}, we present the plots of the regional mean and variance in target number (solid black lines) from the CPHD filter as a function of the radius, for a typical run. For $r = \SI{200}{\meter}$, the mean target number in the region is approximately two with very small variance suggesting that with very high confidence, both targets are covered at $t=51,55$ and $\SI{59}{\second}$. As the radius increases from $r=\SI{1}{\meter}$ (and the circumferences of the regions depart from target $1$), the uncertainty starts increasing until it reaches a local maximum. The behaviour of the variance curves, after the local maximum and until they reach a small steady value, is of concern. In both \figurename~\ref{FigCPHDApproaching}\subref{SubFigCPHDt51}~and~\subref{SubFigCPHDt59}, the local minimum separating the two maximums clearly indicates that target $1$ is contained with high confidence in a circle whose radius equals the value at the mininum (as the mean target number also reaches one at this minimum). When the targets are located at their closest positions, (\figurename~\ref{FigCPHDApproaching}\subref{SubFigCPHDt55}), we cannot identify such regions.
      
      We contrast these results with those obtained after filtering the measurements of an inferior range-bearing sensor which has $\SI{12.5}{\meter}$ and $\SI{2.5}{\degree}$ standard deviations in range and bearing, respectively. The regional variance for this sensor at $t=51,55$ and $\SI{59}{\second}$ (solid red lines in \figurename~\ref{FigCPHDApproaching}\subref{SubFigCPHDt51}--\subref{SubFigCPHDt59}) stays at a high level until the expected target number reaches two, and, in turn, we are unable to select a region that contains only target $1$ with high confidence. In other words, the two targets are not resolved at these time instants.
      
      In \figurename~\ref{FigCPHDApproaching}\subref{SubFigPHDt51}--\subref{SubFigPHDt59}, we present similar results obtained using the PHD filter. We note that the PHD filter performs as well as the CPHD filter in terms of the ability to resolve the two targets in this particular scenario. As a result, the regional variance computed by any of the filters can effectively be used to assess the level of uncertainty in the estimated number of targets in arbitrary regions.
      
  \section{Conclusion} \label{SecConclusion}
    The motivation of this work was to develop multi-object estimators that are able to provide information about the expected number of targets and the uncertainty of the target number in any arbitrary region of the surveillance scene. This level information has never previously been available to operators through track-based multi-target estimators. Providing the regional variance in target number, alongside the regional mean target number, has the potential to give an enhanced picture for surveillance scenarios to address sensor management and resource allocation problems.
    
    Multi-object estimation in a surveillance scene with a challenging environment is the focus of the multi-object paradigm often known as Finite Set Statistics, which leads to filtering algorithms built upon multi-object probability \textit{densities} rather than probability \textit{measures}. However, since such implementations are insufficiently general to represent second-order information about the target number in any arbitrary region, this article adopts a measure-theoretical approach which enables the computation of the regional variance of multi-object estimators. A comprehensive description of the theoretical construction and the practical implementation of the regional mean and variance in target number, in the context of PHD and CPHD filtering, is provided and illustrated on simulated data.
  
  \section*{Acknowledgements}
    This work was supported by the Engineering and Physical Sciences Research Council (EPSRC) Grant number EP/J015180/1 and the MOD University Defence Research Centre on Signal Processing (UDRC). J\'{e}r\'{e}mie Houssineau has a PhD scholarship sponsored by DCNS and a tuition fee scholarship sponsored by Heriot-Watt University.
  \appendices
  \section{Intermediary results} \label{AppIntermediaryResults}
    \begin{property}\textnormal{Normalizing constant (CPHD and PHD updates)\cite{Mahler_RPS_2007}, \cite{Vo_BT_2007}, \cite{Mahler_RPS_2003}} \label{PropCPHDDenominator} \newline
      Under the assumptions given in Theorem \ref{TheoCPHDStatistics}, the denominator of the updated PGFl \eqref{EqMultiBayesPGFlFilter2} becomes
      \begin{equation} \label{EqCPHDDenominator}
	\int L(z_{1:m}|\varphi)P_{\Phi}(d\varphi) \propto \left<\Upsilon^0[\mu_{\Phi},z_{1:m}], \rho_{\Phi}\right>.
      \end{equation}
      \noindent
      Under the assumptions given in Theorem \ref{TheoPHDStatistics}, the denominator of the updated PGFl \eqref{EqMultiBayesPGFlFilter2} becomes
      \begin{equation} \label{EqPHDDenominator}
	\int L(z_{1:m}|\varphi)P_{\Phi}(d\varphi) \propto e^{\mu^{\phi}_{\Phi}(\Xcal)} \prod_{z \in z_{1:m}} (\mu^z_{\Phi}(\Xcal) + \lambda_cc(z)).
      \end{equation}
    \end{property}
    \noindent
    The proof is given in Appendix \ref{AppProofs} (Section \ref{AppPropCPHDDenominator}).
    
  \section{Proofs} \label{AppProofs}
    \subsection{Property \ref{PropCPHDDenominator}} \label{AppPropCPHDDenominator}
      \begin{proof} \label{ProofCPHDDenominator}
	We first focus on the CPHD filter. Using the definition of an i.i.d. process \cite{Vo_BT_2008_2}, the first assumption in Theorem \ref{TheoCPHDStatistics} states that the first moment measure $\mu_{\Phi}$ and the cardinality distribution $\rho_{\Phi}$ are linked by the relation
	\begin{equation} \label{EqIIDCluster}
	  \mu_{\Phi}(\Xcal) = \sum_{n \geqslant 1} n\rho_{\Phi}(n).
	\end{equation}
	They also completely determine the predicted process:
	\begin{equation} \label{EqIIDCluster2}
	  \forall x_{1:n} \in \Xcal^n, P_{\Phi}(dx_{1:n}) = \rho_{\Phi}(n)\prod_{i = 1}^n\frac{\mu_{\Phi}(dx_i)}{\mu_{\Phi}(\Xcal)}.
	\end{equation}
	The remaining assumptions in Theorem \ref{TheoCPHDStatistics} shape the multi-measurement/multi-target likelihood $L$ and yield
	\begin{multline} \label{EqCPHDLikelihood}
	  L(z_{1:m}|x_{1:n}) =
	  \\
	  \sum_{\mathclap{\pi \in \Pi_{m,n}}}~ \pi_{\phi}! \rho_c(\pi_{\phi}) \ds{\prod_{(i,\phi) \in \pi}} c(z_i) \ds{\prod_{(i,j) \in \pi}} P(z_i|x_j) \ds{\prod_{(\phi,j) \in \pi}} P(\phi|x_j),
	\end{multline}
	where:
	\begin{itemize}
	  \item $\Pi_{m,n}$ is the set of all the partitions of indexes $\{i_1,...i_m,j_1,...,j_n\}$ solely composed of tuples of the form $(i_a,j_b)$ (target $x_{j_b}$ is detected and produces measurement $z_{i_a}$), $(\phi, j_b)$ (target $x_{j_b}$ is not detected), or $(i_a, \phi)$ (measurement $z_{i_a}$ is clutter);
	  \item $\pi_{\phi} = \#\{i|(i,\phi) \in \pi\}$ is the number of clutter measurements given by partition $\pi$.
	\end{itemize}
	Note that both the predicted probability measure \eqref{EqIIDCluster2} and the likelihood function \eqref{EqCPHDLikelihood} are \textit{symmetrical} w.r.t. the targets. This property will help simplify the full multi-target Bayes update \eqref{EqMultiBayesPGFlFilter2} to tractable approximations for both PHD and CPHD filters. Substituting \eqref{EqIIDCluster2} into  \eqref{EqMultiBayesPGFlFilter2} gives
	\begin{align}
	  \int L(z_{1:m}|\varphi) &P_{\Phi}(d\varphi) = \nonumber
	  \\
	  &\ds{\sum_{n \geqslant 0}} \rho_{\Phi}(n) \int L(z_{1:m}|x_{1:n}) \prod_{i = 1}^n \frac{\mu_{\Phi}(dx_i)}{\mu_{\Phi}(\Xcal)}. \label{EqCPHDLikelihoodIntegrated}
	\end{align}
	Let us first fix an arbitrary target number $n \in \Nset$ and consider the quantity $\int L(z_{1:m}|x_{1:n}) \prod_{i = 1}^n \frac{\mu_{\Phi}(dx_i)}{\mu_{\Phi}(\Xcal)}$. Since the likelihood is symmetrical w.r.t. the targets, the integration variables $x_{1:n}$ play an identical role and using \eqref{EqCPHDLikelihood} yields
	\begin{multline}
	  \int L(z_{1:m}|x_{1:n}) \prod_{i = 1}^n \frac{\mu_{\Phi}(dx_i)}{\mu_{\Phi}(\Xcal)}
	  \\
	  = \sum_{\mathclap{\pi \in \Pi_{m,n}}}~ \pi_{\emptyset}!\rho_c(\pi_{\emptyset}) \ds{\prod_{(i,\emptyset) \in \pi}} c(z_i) \ds{\prod_{(i,j) \in \pi}} \frac{\mu^{z_i}_{\Phi}(\Xcal)}{\mu_{\Phi}(\Xcal)} \ds{\prod_{(\emptyset,j) \in \pi}} \frac{\mu^{\phi}_{\Phi}(\Xcal)}{\mu_{\Phi}(\Xcal)}. \label{EqCPHDLikelihoodIntegrated2}
	\end{multline}
	Note that, since the targets are identically distributed, measurement/target pairings $(z_i, x_{j_1})$ and $(z_i, x_{j_2})$ are equivalent for integration purpose in \eqref{EqCPHDLikelihoodIntegrated2}. Thus, selecting a partition $\pi \in \Pi_{m,n}$ reduces to the choice of:
	\begin{itemize}
	  \item A number $d$ of detections;
	  \item A collection of $d$ measurements in $z_1,\dots,z_m$;
	  \item An \textit{arbitrary} collection of $d$ detected targets in $x_1,\dots,x_n$.
	\end{itemize}
	Therefore, \eqref{EqCPHDLikelihoodIntegrated2} simplifies as follows:
	\begin{subequations} \label{EqCPHDLikelihoodIntegrated3}
	  \begin{align}
	    &\int L(z_{1:m}|x_{1:n}) \prod_{i = 1}^n \frac{\mu_{\Phi}(dx_i)}{\mu_{\Phi}(\Xcal)} \nonumber
	    \\
	    &\propto \sum_{d = 0}^{\mathclap{\min(m, n)}}~~ \frac{n!(m - d)!}{(n - d)!} \rho_c(m - d) \frac{\mu^{\phi}_{\Phi}(\Xcal)^{n - d}}{\mu_{\Phi}(\Xcal)^n} ~\ds{\sum_{\mathclap{\substack{I \subseteq z_{1:m} \\ |I| = d}}}}~~ \ds{\prod_{z \in I}} \frac{\mu^z_{\Phi}(\Xcal)}{c(z)} \label{EqCPHDLikelihoodIntegrated3A}
	    \\
	    &\propto \sum_{d = 0}^{\mathclap{\min(m, n)}}~~ \frac{n!(m - d)!}{(n - d)!} \rho_c(m - d) \frac{\mu^{\phi}_{\Phi}(\Xcal)^{n - d}}{\mu_{\Phi}(\Xcal)^n} e_d(z_{1:m}) \label{EqCPHDLikelihoodIntegrated3B}
	    \\
	    &\propto \Upsilon^0[\mu_{\Phi},z_{1:m}](n), \label{EqCPHDLikelihoodIntegrated3C}      
	  \end{align}
	\end{subequations}
	using the $\Upsilon$ function defined in \eqref{EqVoNotation1}. The multiplying constant in \eqref{EqCPHDLikelihoodIntegrated}, found to be $\prod_{z \in z_{1:m}} c(z)$, will appear as well in the expression of the numerator of the updated PGFl \eqref{EqMultiBayesPGFlFilter2} developed in Appendix \ref{AppIntermediaryResults} in Section \ref{AppLemCPHDFirstMoment} and \ref{AppLemCPHDSecondMoment} will be omitted from now on. Finally, substituting \eqref{EqCPHDLikelihoodIntegrated3B} in \eqref{EqCPHDLikelihoodIntegrated} yields the result \eqref{EqCPHDDenominator}.\newline
	\\
	We now move to the PHD filter. Since a Poisson process is a specific case of a i.i.d. process, we start from the CPHD result \eqref{EqCPHDDenominator} with the \textit{additional} assumptions that:
	\begin{enumerate}
	  \item The predicted process is Poisson: $\rho_{\Phi}(n) = e^{-\mu_{\Phi}(\Xcal)}\frac{\mu_{\Phi}(\Xcal)^n}{n!}$;
	  \item The clutter process is Poisson: $\rho_c(n) = e^{-\lambda_c}\frac{\lambda_c^n}{n!}$ and $\lambda_c = \sum_{n \geqslant 0} n \rho_c(n)$.
	\end{enumerate}
	We may write:
	\begin{subequations} \label{EqPHDDenominatorProof}
	  \begin{align}
	    &\int L(z_{1:m}|\varphi)P_{\Phi}(d\varphi) \propto \left<\Upsilon^0[\mu_{\Phi},z_{1:m}], \rho_{\Phi}\right> \label{EqPHDDenominatorA}
	    \\
	    &\propto \ds{\sum_{n \geqslant 0}} \rho_{\Phi}(n) \sum_{d = 0}^{\mathclap{\min(m, n)}}~~ \frac{n!(m - d)!}{(n - d)!} \rho_c(m - d) \frac{\mu^{\phi}_{\Phi}(\Xcal)^{n - d}}{\mu_{\Phi}(\Xcal)^n} e_d(z_{1:m}) \label{EqPHDDenominatorB}
	    \\
	    &\propto \ds{\sum_{n \geqslant 0}} \sum_{d = 0}^{\min(m, n)} \frac{1}{(n - d)!} \lambda_c^{m - d} \mu^{\phi}_{\Phi}(\Xcal)^{n - d} e_d(z_{1:m}) \label{EqPHDDenominatorC}
	    \\
	    &\propto \sum_{d = 0}^{m} \left(\ds{\sum_{n \geqslant d}} \frac{\mu^{\phi}_{\Phi}(\Xcal)^{n - d}}{(n - d)!}\right) \lambda_c^{m - d} \ds{\sum_{\substack{I \subseteq z_{1:m} \\ |I| = d}}} \ds{\prod_{z \in I}} \frac{\mu^z_{\Phi}(\Xcal)}{c(z)} \label{EqPHDDenominatorD}
	    \\
	    &\propto e^{\mu^{\phi}_{\Phi}(\Xcal)} \sum_{d = 0}^{m} \ds{\sum_{\substack{I \subseteq z_{1:m} \\ |I| = d}}}~\ds{\prod_{z \in I}} \mu^z_{\Phi}(\Xcal)~\ds{\prod_{z \notin I}} \lambda_cc(z) \label{EqPHDDenominatorE}
	    \\
	    &\propto e^{\mu^{\phi}_{\Phi}(\Xcal)} \prod_{z \in z_{1:m}} (\mu^z_{\Phi}(\Xcal) + \lambda_cc(z)), \label{EqPHDDenominatorF}
	  \end{align}
	\end{subequations}
	where \eqref{EqPHDDenominatorF} is the factorised form of \eqref{EqPHDDenominatorE}.
      \end{proof}
      
    \subsection{Lemma \ref{LemCPHDFirstMoment}} \label{AppLemCPHDFirstMoment} 
      \begin{proof} \label{ProofCPHDFirstMoment}
	Using \eqref{EqFirstMomentComputation}, the first moment measure $\mu_{\Phi_{+}}$ in some $B \in \Bbold_{\Xcal}$ is retrieved from the first order differential \cite{Clark_DE_2012_2} of the updated PGFl \eqref{EqMultiBayesPGFlFilter2}:
	\begin{subequations} \label{EqFirstMoment1}
	  \begin{align}
	    &\mu_{\Phi_{+}}(B) = \left.\delta(\Gcal_{\Phi_{+}}[h]; 1_{B})\right|_{h = 1} \label{EqFirstMoment1A}
	    \\
	    &= \frac{\ds{\sum_{n \geqslant 0}} \int \left.\delta\left(\prod_{i = 1}^n h(x_i) ; 1_{B}\right)\right|_{h = 1} L(z_{1:m}|x_{1:n}) P_{\Phi}(dx_{1:n})}{\ds{\sum_{n \geqslant 0}} \int L(z_{1:m}|x_{1:n}) P_{\Phi}(dx_{1:n})}. \label{EqFirstMoment1B}
	  \end{align}
	\end{subequations}
	The expression of the denominator in \eqref{EqFirstMoment1B} is detailed separately in Property \ref{PropCPHDDenominator} (Section \ref{AppIntermediaryResults}). Using Corollary 1 in \cite{Clark_DE_2012_2}, the numerator expands as follows:
	\begin{multline} \label{EqFirstMoment1.2}
	  \ds{\sum_{n \geqslant 0}} \int \left.\delta\left(\prod_{i = 1}^n h(x_i) ; 1_{B}\right)\right|_{h = 1} L(z_{1:m}|x_{1:n}) P_{\Phi}(dx_{1:n}) 
	  \\
	  = \ds{\sum_{n \geqslant 1}} \int \left(\sum_{1 \leqslant j \leqslant n} \prod_{i = 1}^n \mu^j_i(x_i)\right) L(z_{1:m}|x_{1:n}) P_{\Phi}(dx_{1:n}),
	\end{multline}
	where $\mu^j_i = 1_{B}$ if $i = j$, $\mu^j_i = 1$ otherwise. Thus:
	\begin{multline} \label{EqFirstMoment1.3}
	  \ds{\sum_{n \geqslant 0}} \int \left.\delta\left(\prod_{i = 1}^n h(x_i) ; 1_{dy}\right)\right|_{h = 1} L(z_{1:m}|x_{1:n}) P_{\Phi}(dx_{1:n})
	  \\
	  = \ds{\sum_{n \geqslant 1}} \int \sum_{1 \leqslant j \leqslant n} 1_{B}(x_j) L(z_{1:m}|x_{1:n}) P_{\Phi}(dx_{1:n}).
	\end{multline}
	\noindent
	As seen in \eqref{EqIIDCluster2} and \eqref{EqCPHDLikelihood} in the construction of the denominator (proof of Property \ref{PropCPHDDenominator} in Section \ref{ProofCPHDDenominator}), $L(z_{1:m}|x_{1:n})$ and $P_{\Phi}(dx_{1:n})$ are \textit{symmetrical} w.r.t. to the targets in the specific case of the CPHD filter. Thus \eqref{EqFirstMoment1.3} simplifies as follows:
	\begin{subequations} \label{EqFirstMoment2}
	  \begin{align}
	    &\ds{\sum_{n \geqslant 0}} \int \left.\delta\left(\prod_{i = 1}^n h(x_i) ; 1_{dy}\right)\right|_{h = 1} L(z_{1:m}|x_{1:n}) P_{\Phi}(dx_{1:n}) \nonumber
	    \\
	    &= \ds{\sum_{n \geqslant 1}} n \int 1_{B}(x) L(z_{1:m}|x_{1:n-1},x) P_{\Phi}(dx_{1:n-1},dx) \label{EqFirstMoment2A}
	    \\
	    &= \ds{\sum_{n \geqslant 1}} \frac{n\rho_{\Phi}(n)}{\mu_{\Phi}(\Xcal)} \int 1_{B}(x) L(z_{1:m}|x_{1:n-1}, x) \nonumber
	    \\
	    &\quad\quad\quad\quad\quad\quad\quad\quad\quad\quad\quad\times \mu_{\Phi}(dx) \prod_{i = 1}^{n - 1} \frac{\mu_{\Phi}(dx_i)}{\mu_{\Phi}(\Xcal)}. \label{EqFirstMoment2B}
	  \end{align}
	\end{subequations}
	Now, considering the expression of the likelihood \eqref{EqCPHDLikelihood}, the likelihood term in \eqref{EqFirstMoment2A} can be split following partitions where target $x$ is not detected and those where it is detected and produces a particular measurement $z \in z_{1:m}$, i.e.
	\begin{multline} \label{EqLikelihoodSplit}
	  L(z_{1:m}|x_{1:n-1}, x) = 
	  \\
	  P(\phi|x)L(z_{1:m}|x_{1:n-1}) + \sum_{z \in z_{1:m}}P(z|x)L(z_{1:m} \setminus z|x_{1:n-1}).
	\end{multline}
	Substituting \eqref{EqLikelihoodSplit} in \eqref{EqFirstMoment2B}, then substituting the result in the expression of the first moment measure \eqref{EqFirstMoment1B} finally yields
	\begin{multline} \label{EqFirstMoment3}
	  \mu_{\Phi_{+}}(B) = \left(\int 1_B(x) P(\phi|x) \mu_{\Phi}(dx)\right) \ell_1(\phi)
	  \\
	  + \sum_{\mathclap{z \in z_{1:m}}}~ \frac{\int 1_B(x) P(z|x) \mu_{\Phi}(dx)}{c(z)}\ell_1(z),
	\end{multline}
	where the corrector terms $\ell_1(\phi)$ and $L(z)$, following a similar development as in the proof of Property \ref{PropCPHDDenominator}, are found to be
	\begin{subequations} \label{EqFirstMomentCorrector1}
	  \begin{align}
	    \ell_1(\phi) &= \frac{\ds{\sum_{n \geqslant 1}} \frac{n\rho_{\Phi}(n)}{\mu_{\Phi}(\Xcal)} \int L(z_{1:m}|x_{1:n-1}) \prod_{i = 1}^{n - 1} \frac{\mu_{\Phi}(dx_i)}{\mu_{\Phi}(\Xcal)}}{\ds{\sum_{n \geqslant 0}} \int L(z_{1:m}|x_{1:n}) P_{\Phi}(dx_{1:n})} \label{EqFirstMomentCorrector1A}
	    \\
	    &= \frac{\left<\Upsilon^1[\mu_{\Phi},z_{1:m}], \rho_{\Phi}\right>}{\left<\Upsilon^0[\mu_{\Phi},z_{1:m}], \rho_{\Phi}\right>}, \label{EqFirstMomentCorrector1B}
	  \end{align}
	\end{subequations}
	and:
	\begin{subequations} \label{EqFirstMomentCorrector2}
	  \begin{align}
	    \ell_1(z) &= \frac{c(z)\ds{\sum_{n \geqslant 1}} \frac{n\rho_{\Phi}(n)}{\mu_{\Phi}(\Xcal)} \int L(z_{1:m} \setminus z|x_{1:n-1}) \prod_{i = 1}^{n - 1} \frac{\mu_{\Phi}(dx_i)}{\mu_{\Phi}(\Xcal)}}{\ds{\sum_{n \geqslant 0}} \int L(z_{1:m}|x_{1:n}) P_{\Phi}(dx_{1:n})} \label{EqFirstMomentCorrector2A}
	    \\
	    &= \frac{\left<\Upsilon^1[\mu_{\Phi},z_{1:m} \setminus z], \rho_{\Phi}\right>}{\left<\Upsilon^0[\mu_{\Phi},z_{1:m}], \rho_{\Phi}\right>}. \label{EqFirstMomentCorrector2B}
	  \end{align}
	\end{subequations}
      \end{proof}
      
    \subsection{Corollary \ref{CorPHDFirstMoment}} \label{AppCorPHDFirstMoment}
      \begin{proof} \label{ProofPHDFirstMoment}
	Just as the Poisson assumption simplified the expression of $\Upsilon^0$ as shown in the development \eqref{EqPHDDenominatorProof}, it simplifies the expression of $\Upsilon^1$:
	\begin{align}
	  \left<\Upsilon^1[\mu_{\Phi},z_{1:m}],\rho_{\Phi}\right> &\propto e^{\mu^{\phi}_{\Phi}(\Xcal)} \prod_{\mathclap{z \in z_{1:m}}}~ (\mu^z_{\Phi}(\Xcal) + \lambda_cc(z)), \label{EqPHDNumerator1}
	  \\
	  \left<\Upsilon^1[\mu_{\Phi},z_{1:m} \setminus z],\rho_{\Phi}\right> &\propto c(z) e^{\mu^{\phi}_{\Phi}(\Xcal)} \prod_{\mathclap{z' \in z_{1:m} \setminus z}}~ (\mu^{z'}_{\Phi}(\Xcal) + \lambda_cc(z')). \label{EqPHDNumerator2}
	\end{align}
	Then, substituting the simplified expressions of $\Upsilon^0$ \eqref{EqPHDDenominatorF} and $\Upsilon^1$ \eqref{EqPHDNumerator1}, \eqref{EqPHDNumerator2} in the first moment measure of the CPHD filter \eqref{EqCPHDFirstMoment} yields the result for the PHD filter \eqref{EqPHDFirstMoment}.
      \end{proof}
      
    \subsection{Lemma \ref{LemCPHDSecondMoment}} \label{AppLemCPHDSecondMoment}
      \begin{proof} \label{ProofCPHDSecondMoment}
	Using \eqref{EqSecondMomentComputation}, the updated second moment measure $\mu^{(2)}_{\Phi_{+}}$ in some regions $B,~B' \in \Bbold_{\Xcal}$ is retrieved from the second-order differential \cite{Clark_DE_2012_2} of the updated Laplace functional \eqref{EqMultiBayesLaplaceFilter2}:
	\begin{subequations} \label{EqSecondMoment1}
	  \begin{align}
	    &\mu^{(2)}_{\Phi_{+}}(B, B') = \left.\delta(\Lcal_{\Phi_{+}}[f]; 1_{B}, 1_{B'})\right|_{f = 0} \label{EqSecondMoment1A}
	    \\
	    &= \frac{\ds{\sum_{n \geqslant 0}} \int \left.\delta^2(e^{-\sum f(x_i)}; 1_{B}, 1_{B'})\right|_{f = 0} L(z_{1:m}|x_{1:n}) P_{\Phi}(dx_{1:n})}{\ds{\sum_{n \geqslant 0}} \int L(z_{1:m}|x_{1:n}) P_{\Phi}(dx_{1:n})}. \label{EqSecondMoment1B}
	  \end{align}
	\end{subequations}
	The second-order differential in \eqref{EqSecondMoment1B} is found to be
	\begin{multline} \label{EqSecondOrderDerivative}
	  \left.\delta^2(e^{-\sum_{i = 1}^n f(x_i)}; 1_{B}, 1_{B'})\right|_{f = 0}
	  \\
	  = \ds{\sum_{1 \leqslant j \leqslant n}} 1_{B \cap B'}(x_j) + \ds{\sideset{}{^{\neq}}\sum_{1 \leqslant j_1, j_2 \leqslant n}} 1_{B}(x_{j_1})1_{B'}(x_{j_2}), 
	\end{multline}
	the proof being given in Appendix \ref{AppProofs} (Section \ref{AppSecondOrderDerivation}). Substituting \eqref{EqSecondOrderDerivative} in the numerator of \eqref{EqSecondMoment1} gives
	\begin{align}
	  &\ds{\sum_{n \geqslant 0}} \int \left.\delta^2(e^{-\sum f(x_i)}; 1_{B}, 1_{B'})\right|_{f = 0} L(z_{1:m}|x_{1:n}) P_{\Phi}(dx_{1:n}) \nonumber
	  \\
	  &= \ds{\sum_{n \geqslant 1}} \int \left(~\ds{\sum_{\mathclap{1 \leqslant j \leqslant n}}}~ 1_{B \cap B'}(x_j)\right) L(z_{1:m}|x_{1:n}) P_{\Phi}(dx_{1:n}) \nonumber
	  \\
	  &+ \ds{\sum_{n \geqslant 2}} \int \left(~~\ds{\sideset{}{^{\neq}}\sum_{\mathclap{1 \leqslant j_1, j_2 \leqslant n}}}~ 1_{B}(x_{j_1})1_{B'}(x_{j_2})\right) L(z_{1:m}|x_{1:n}) P_{\Phi}(dx_{1:n}). \label{EqSecondMoment1.2}
	\end{align}
	\noindent
	Once again, the symmetry of $L(z_{1:m}|x_{1:n})$ and $P_{\Phi}(dx_{1:n})$ w.r.t. to the targets in the case of the CPHD filter (see \eqref{EqIIDCluster2} and \eqref{EqCPHDLikelihood}) allows the simplification of \eqref{EqSecondMoment1.2}. We have:
	\begin{subequations} \label{EqSecondMoment2}
	  \begin{align}
	    &\ds{\sum_{n \geqslant 0}} \int \left.\delta^2(e^{-\sum f(x_i)}; 1_{B}, 1_{B'})\right|_{f = 0} L(z_{1:m}|x_{1:n}) P_{\Phi}(dx_{1:n}) \nonumber
	    \\
	    &= \ds{\sum_{n \geqslant 1}} n \int 1_{B \cap B'}(x) L(z_{1:m}|x_{1:n-1}, x) P_{\Phi}(dx_{1:n-1}, dx) \nonumber
	    \\
	    &+ \ds{\sum_{n \geqslant 2}} n(n - 1) \int 1_{B}(x) 1_{B'}(x') L(z_{1:m}|x_{1:n-2}, x, x') \nonumber
	    \\
	    &\quad\quad\quad\quad\quad\quad\quad\quad\quad\quad\quad\times P_{\Phi}(dx_{1:n-2}, dx, dx') \label{EqSecondMoment2A}
	    \\
	    &= \ds{\sum_{n \geqslant 1}} \frac{n\rho_{\Phi}(n)}{\mu_{\Phi}(\Xcal)} \int 1_{B \cap B'}(x) L(z_{1:m}|x_{1:n-1}, x) \nonumber
	    \\
	    &\quad\quad\quad\quad\quad\quad\quad\quad\quad\quad\quad\times \mu_{\Phi}(dx) \prod_{i = 1}^{n - 1} \frac{\mu_{\Phi}(dx_i)}{\mu_{\Phi}(\Xcal)} \nonumber
	    \\
	    &+ \ds{\sum_{n \geqslant 2}} \frac{n(n - 1)\rho_{\Phi}(n)}{\mu_{\Phi}(\Xcal)^2} \int 1_{B}(x) 1_{B'}(x') L(z_{1:m}|x_{1:n-2}, x, x') \nonumber
	    \\
	    &\quad\quad\quad\quad\quad\quad\quad\quad\times \mu_{\Phi}(dx)\mu_{\Phi}(dx') \prod_{i = 1}^{n - 2} \frac{\mu_{\Phi}(dx_i)}{\mu_{\Phi}(\Xcal)}. \label{EqSecondMoment2B}
	  \end{align}
	\end{subequations}
	The first likelihood term in \eqref{EqSecondMoment2B}, just as in the proof of Lemma \ref{LemCPHDFirstMoment}, expands following \eqref{EqLikelihoodSplit}. Now, considering the general expression of the likelihood \eqref{EqCPHDLikelihood}, the second likelihood term in \eqref{EqSecondMoment2B} can be split following partitions where none of the targets $x$, $x'$ are detected, those where only one is detected and those where both are detected. That is:
	\begin{align} 
	  &L(z_{1:m}|x_{1:n-2}, x, x') \nonumber
	  \\
	  &= P(\phi|x)P(\phi|x')L(z_{1:m}|x_{1:n-2}) \nonumber
	  \\
	  &+ P(\phi|x)\sum_{z \in z_{1:m}}P(z|x')L(z_{1:m} \setminus z|x_{1:n-2}) \nonumber
	  \\
	  &+ P(\phi|x')\sum_{z \in z_{1:m}}P(z|x)L(z_{1:m} \setminus z|x_{1:n-2}) \nonumber
	  \\
	  &+ \ds{\sideset{}{^{\neq}}\sum_{z, z' \in z_{1:m}}}P(z|x)P(z'|x')L(z_{1:m} \setminus \{z, z'\}|x_{1:n-2}). \label{EqLikelihoodSplit2}
	\end{align}
	Substituting \eqref{EqLikelihoodSplit2} and \eqref{EqLikelihoodSplit} in \eqref{EqSecondMoment2B}, then substituting the result in the expression of the second moment measure \eqref{EqSecondMoment1B} finally yields
	\begin{align}
	  &\mu^{(2)}_{\Phi_{+}}(B, B') = \int 1_{B \cap B'}(x) \mu_{\Phi_{+}}(dx) \nonumber
	  \\
	  &+ \int 1_{B}(x) P(\phi|x)\mu_{\Phi}(dx) \int 1_{B'}(x) P(\phi|x)\mu_{\Phi}(dx) \times \ell_2(\phi) \nonumber
	  \\
	  &+ \int 1_{B}(x) P(\phi|x)\mu_{\Phi}(dx) \sum_{z \in z_{1:m}} \frac{\int 1_{B'}(x) P(z|x)\mu_{\Phi}(dx)}{c(z)}\ell_2(z) \nonumber
	  \\
	  &+ \int 1_{B'}(x) P(\phi|x)\mu_{\Phi}(dx) \sum_{z \in z_{1:m}} \frac{\int 1_{B}(x) P(z|x)\mu_{\Phi}(dx)}{c(z)}\ell_2(z) \nonumber
	  \\
	  &+ \ds{\sideset{}{^{\neq}}\sum_{\mathclap{z, z' \in z_{1:m}}}}\frac{\int 1_{B}(x) P(z|x)\mu_{\Phi}(dx)}{c(z)}\frac{\int 1_{B'}(x) P(z'|x)\mu_{\Phi}(dx)}{c(z')}\ell_2(z, z'), \label{EqSecondMoment3}
	\end{align}
	where the corrector terms $\ell_2(\phi)$, $\ell_2(z)$, and $\ell_2(z, z')$, following a similar development as shown in the proofs of Property \ref{PropCPHDDenominator} (Section \ref{ProofCPHDDenominator}) and \ref{LemCPHDFirstMoment} (Section \ref{ProofCPHDFirstMoment}), are as defined by \eqref{EqCPHDSecondMomentCorrector}.
      \end{proof}
  
    \subsection{Expansion of $\left.\delta^2(e^{-\sum_{i = 1}^n f(x_i)}; 1_{B}, 1_{B'})\right|_{f = 0}$} \label{AppSecondOrderDerivation}
      \begin{proof}
	Expanding the exponential gives
	\begin{align}
	  &\left.\delta^2(e^{-\sum_{i = 1}^n f(x_i)}; 1_{B}, 1_{B'})\right|_{f = 0} \nonumber
	  \\
	  &= \ds{\sum_{p \geqslant 0}} \frac{(-1)^p}{p!} \left.\delta^2\left(\left(\sum_{i = 1}^n f(x_i)\right)^p; 1_{B}, 1_{B'}\right)\right|_{f = 0} \nonumber
	  \\
	  &= \ds{\sum_{p \geqslant 0}} \frac{(-1)^p}{p!} \ds{~~~\sum_{\mathclap{\substack{\\ p_1 + \dots + p_n = p}}}~~~} {p \choose p_{1:n}}\left.\delta^2\left(\prod_{i = 1}^n f(x_i)^{p_i}; 1_{B}, 1_{B'}\right)\right|_{f = 0}, \nonumber
	\end{align}
	where ${p \choose p_{1:n}}$ is the multinomial
	\begin{equation}
	  {p \choose p_{1:n}} = {p \choose p_1,\dots,p_n} = \frac{p!}{p_1!\dots p_n!}.
	\end{equation}
	Then, using Corollary 1 in \cite{Clark_DE_2012_2} yields
	\begin{align}
	  &\left.\delta^2\left(\prod_{i = 1}^n f(x_i)^{p_i}; 1_{B}, 1_{B'}\right)\right|_{f = 0} \nonumber
	  \\
	  &= \ds{\sum_{p_j \geqslant 2}}2{p_j \choose 2}1_{B}(x_j)1_{B'}(x_j)0^{\sum p_i - 2} \nonumber
	  \\
	  &+ \ds{\sum_{\substack{p_{j_1}, p_{j_2} \geqslant 1 \\ j_1 \neq j_2}}}{p_{j_1} \choose 1}{p_{j_2} \choose 1} 1_{B}(x_{j_1})1_{B'}(x_{j_2})0^{\sum p_i - 2}. \nonumber
	\end{align}
	Thus, it follows that
	\begin{align}
	  &\ds{\sum_{p \geqslant 0}} \frac{(-1)^p}{p!} \ds{~~~\sum_{\mathclap{\substack{\\ p_1 + \dots + p_n = p}}}~~~} {p \choose p_{1:n}}\left.\delta^2\left(\prod_{i = 1}^n f(x_i)^{p_i}; 1_{B}, 1_{B'}\right)\right|_{f = 0} \nonumber
	  \\
	  &= \frac{(-1)^2}{2!} \ds{~~~\sum_{\mathclap{\substack{\\ p_1 + \dots + p_n = 2 \\ \exists j | p_j \geqslant 2}}}~~~} 2{2 \choose p_{1:n}}{p_j \choose 2} 1_{B \cap B'}(x_j) \nonumber
	  \\
	  &+ \frac{(-1)^2}{2!} \ds{~~~\sum_{\mathclap{\substack{\\ p_1 + \dots + p_n = 2 \\ \exists j_1 \neq j_2 | p_{j_1}, p_{j_2} \geqslant 1}}}~~~} {2 \choose p_{1:n}} {p_{j_1} \choose 1}{p_{j_2} \choose 1} 1_{B}(x_{j_1})1_{B'}(x_{j_2}) \nonumber
	  \\
	  &= \frac{1}{2} \ds{\sum_{1 \leqslant j \leqslant n}} 2{2 \choose 2, 0}{2 \choose 2} 1_{B \cap B'}(x_j) \nonumber
	  \\
	  &+ \frac{1}{2}~\ds{\sideset{}{^{\neq}}\sum_{1 \leqslant j_1, j_2 \leqslant n}} {2 \choose 1, 1} {1 \choose 1}{1 \choose 1} 1_{B}(x_{j_1})1_{B'}(x_{j_2}) \nonumber
	  \\
	  &= \ds{\sum_{1 \leqslant j \leqslant n}}1_{B \cap B'}(x_j) + \ds{\sideset{}{^{\neq}}\sum_{1 \leqslant j_1, j_2 \leqslant n}} 1_{B}(x_{j_1})1_{B'}(x_{j_2}).\nonumber
	\end{align}
      \end{proof}

    \subsection{Corollary \ref{CorPHDSecondMoment}} \label{AppCorPHDSecondMoment}
      \begin{proof}
	Just as the Poisson assumption simplified the expression of $\Upsilon^0$ as shown in the development \eqref{EqPHDDenominatorProof}, it simplifies the expression of $\Upsilon^2$:
	\begin{align}
	  &\left<\Upsilon^2[\mu_{\Phi},z_{1:m}], \rho_{\Phi}\right> \propto e^{\mu^{\phi}_{\Phi}(\Xcal)} ~\prod_{\mathclap{z \in z_{1:m}}}~ (\mu^z_{\Phi}(\Xcal) + \lambda_cc(z)), \label{EqPHDNumerator3}
	  \\
	  &\left<\Upsilon^2[\mu_{\Phi},z_{1:m} \setminus z], \rho_{\Phi}\right> \nonumber
	  \\
	  &\quad\quad\quad \propto c(z) e^{\mu^{\phi}_{\Phi}(\Xcal)} ~\prod_{\mathclap{z' \in z_{1:m} \setminus z}}~ (\mu^{z'}_{\Phi}(\Xcal) + \lambda_cc(z')), \label{EqPHDNumerator4}
	  \\
	  &\left<\Upsilon^2[\mu_{\Phi},z_{1:m} \setminus \{z,z'\}], \rho_{\Phi}\right> \nonumber
	  \\
	  &\quad\quad\quad \propto c(z)c(z') e^{\mu^{\phi}_{\Phi}(\Xcal)} ~\prod_{\mathclap{z'' \in z_{1:m} \setminus \{z, z'\}}}~ (\mu^{z''}_{\Phi}(\Xcal) + \lambda_cc(z'')). \label{EqPHDNumerator5}
	\end{align}
	Then, substituting the simplified expressions of $\Upsilon^0$ \eqref{EqPHDDenominatorF}, $\Upsilon^1$ \eqref{EqPHDNumerator1}, \eqref{EqPHDNumerator2}, and $\Upsilon^2$ \eqref{EqPHDNumerator3}, \eqref{EqPHDNumerator4}, \eqref{EqPHDNumerator5} in the second moment measure of the CPHD filter \eqref{EqCPHDSecondMoment} yields the result for the PHD filter \eqref{EqPHDSecondMoment}.
      \end{proof}
      
    \subsection{Theorems \ref{TheoCPHDStatistics} and \ref{TheoPHDStatistics}} \label{AppTheoCPHDPHDStatistics}
      \begin{proof} \label{ProofCPHDStatistics}
	The first order statistic $\mu_{\Phi_{+}}(B)$ is given by Lemma \ref{LemCPHDFirstMoment}. Following the definition of the variance \eqref{EqVariance}, the second-order statistic $\var_{\Phi_{+}}(B)$ is the second moment measure $\mu^{(2)}_{\Phi_{+}}(B, B')$ (Lemma \ref{LemCPHDSecondMoment}) with $B'= B$, from which $\left[\mu_{\Phi_{+}}(B)\right]^2$ is substracted. This concludes the proof of Theorem \ref{TheoCPHDStatistics}.\newline
	\\
	The proof of Theorem \ref{TheoPHDStatistics} is identical, except that Corollaries \ref{CorPHDFirstMoment} and \ref{CorPHDSecondMoment} are used instead of Lemmas \ref{LemCPHDFirstMoment} and \ref{LemCPHDSecondMoment}.
      \end{proof}
      
  \section{Algorithms} \label{AppAlgorithms}
    \begin{algorithm}[H]
      \begin{algorithmic}
	\State \emph{Input}
	  \State Predicted intensity: $\{w^{(i)}, x^{(i)}\}_{i = 1}^J$
	  \State Cardinality distribution: $\{\rho(n)\}_{n = 0}^{n_{\max}}$
	  \State Current measurements: $z_{1:m}$
	  \State Maximum cardinality: $n_{\max}$
	\\
	\State \emph{Missed detection and measurement terms}
	  \For{$1 \leqslant i \leqslant J$}
	    \State $w^{(i), \phi} \leftarrow P(\phi|x^{(i)})w^{(i)}$
	    \For{$z_k \in z_{1:m}$}
	      \State $w^{(i), z_k} \leftarrow P(z_k|x^{(i)})w^{(i)}$
	    \EndFor
	  \EndFor
	\State Compute global missed detection term
	  \State $\mu^{\phi}_{\Phi}(\Xcal) \leftarrow \sum_{i = 1}^J w^{(i), \phi}$
	\State Compute global measurement terms
	  \For{$z_k \in z_{1:m}$}
	    \State $\mu^{z_k}_{\Phi}(\Xcal) \leftarrow \sum_{i = 1}^J w^{(i), z_k}$
	  \EndFor
	\\
	\State \emph{Corrector terms}
	  \State Compute $e_d(z_{1:m})$ using \eqref{EqElementarySymmetricFunction}
	  \For{$0 \leqslant n \leqslant n_{\max}$}
	    \State Compute $\{\Upsilon^0, \Upsilon^1, \Upsilon^2\}[\mu_{\Phi},z_{1:m}](n)$ using \eqref{EqVoNotation1}
	  \EndFor
	  \State Compute $\ell_1(\phi)$ using \eqref{EqCPHDFirstMomentCorrector} and $\ell_2(\phi)$ using \eqref{EqCPHDSecondMomentCorrector}
	  \For{$z_k \in z_{1:m}$}
	    \State Compute $e_d(z_{1:m} \setminus z_k)$ using \eqref{EqElementarySymmetricFunction}
	    \For{$0 \leqslant n \leqslant n_{\max}$}
	      \State Compute $\{\Upsilon^1, \Upsilon^2\}[\mu_{\Phi},z_{1:m} \setminus z_k](n)$ using \eqref{EqVoNotation1}
	    \EndFor
	    \State Compute $\ell_1(z_k)$ using \eqref{EqCPHDFirstMomentCorrector} and $\ell_2(z_k)$ using \eqref{EqCPHDSecondMomentCorrector}
	      \For{$z_l \in z_{1:m},~l > k$}
		\State Compute $e_d(z_{1:m} \setminus \{z_k, z_l\})$ using \eqref{EqElementarySymmetricFunction}
		\For{$0 \leqslant n \leqslant n_{\max}$}
		  \State Compute $\Upsilon^2[\mu_{\Phi},z_{1:m} \setminus \{z_k, z_l\}](n)$ using \eqref{EqVoNotation1}
		\EndFor
		\State Compute $\ell_2(z_k, z_l)$ using \eqref{EqCPHDSecondMomentCorrector}
	      \EndFor
	  \EndFor
	  \\
	\State \emph{Data update}
	  \State Update cardinality distribution
	    \For{$0 \leqslant n \leqslant n_{\max}$}
	      \State $\rho_{+}(n) \leftarrow \frac{\Upsilon^0[\mu_{\Phi},z_{1:m}](n)\rho(n)}{\sum_{n' = 0}^{n_{\max}}\Upsilon^0[\mu_{\Phi},z_{1:m}](n')}$
	    \EndFor
	  \State Update intensity
	    \For{$1 \leqslant i \leqslant J$}
	      \State $w^{(i)}_{+} \leftarrow w^{(i), \phi}\ell_1(\phi) + \sum_{z_k \in z_{1:m}} \frac{w^{(i), z_k}}{c(z_k)}\ell_1(z_k)$
	    \EndFor
	\\
	\algstore{bkbreak}
      \end{algorithmic}
      \caption{CPHD filter with variance: data update (adapted from \cite{Vo_BT_2008_2}) and information statistics\label{AlgCPHD}}
    \end{algorithm}
    \addtocounter{algorithm}{-1}
    \begin{algorithm}[H]
      \begin{algorithmic}
	\algrestore{bkbreak}
	\State \emph{Regional terms}
	  \State $\mu^{\phi}_{\Phi}(B) \leftarrow \sum_{x^{(i)} \in B} w^{(i), \phi}$
	  \For{$z_k \in z_{1:m}$}
	    \State $\mu^{z_k}_{\Phi}(B) \leftarrow  \sum_{x^{(i)} \in B} w^{(i), z_k}$
	  \EndFor
	\\
	\State \emph{Mean target number}
	  \State $\mu_{\Phi_{+}}(B) \simeq \mu^{\phi}_{\Phi}(B)\ell_1(\phi) + \sum_{z_k \in z_{1:m}} \frac{\mu^{z_k}_{\Phi}(B)}{c(z_k)}\ell_1(z_k)$
	\\
	\State \emph{Variance in target number}  
	  \State $var_{\Phi_{+}}(B) \simeq \mu_{\Phi_{+}}(B) + \mu^{\phi}_{\Phi}(B)^2\left[\ell_2(\phi) - \ell_1(\phi)^2\right]$
	  \State $\quad\quad\quad\quad+ 2\mu^{\phi}_{\Phi}(B) \sum_{k = 1}^m \frac{\mu^{z_k}_{\Phi}(B)}{c(z_k)} \left[\ell_2(z_k) - \ell_1(z_k)\ell_1(\phi)\right]$
	  \State $\quad\quad\quad\quad+ 2\ds{\sum_{\mathclap{1 \leqslant k < l \leqslant m}}}~ \frac{\mu^{z_k}_{\Phi}(B)}{c(z_k)} \frac{\mu^{z_l}_{\Phi}(B)}{c(z_l)} \left[\ell_2(z_k,z_l) - \ell_1(z_k)\ell_1(z_l)\right]$
	  \State $\quad\quad\quad\quad- \sum_{k = 1}^m \left(\frac{\mu^{z_k}_{\Phi}(B)}{c(z_k)}\ell_1(z_k)\right)^2$
      \end{algorithmic}
      \caption{CPHD filter with variance (cont.)}
    \end{algorithm}
    
    \begin{algorithm}[H]
      \begin{algorithmic}
	\State \emph{Input}
	  \State Predicted intensity: $\{w^{(i)}, x^{(i)}\}_{i = 1}^J$
	  \State Current measurements: $z_{1:m}$
	\\
	\State \emph{Missed detection and measurement terms}
	  \For{$1 \leqslant i \leqslant J$}
	    \State Compute missed detection term
	    \State $w^{(i), \phi} \leftarrow P(\phi|x^{(i)})w^{(i)}$
	    \State Compute measurement terms
	    \For{$z_k \in z_{1:m}$}
	      \State $\hat{w}^{(i), z_k} \leftarrow P(z_k|x^{(i)})w^{(i)}$
	    \EndFor
	  \EndFor
	\\
	\State \emph{Data update}
	  \For{$1 \leqslant i \leqslant J$}
	    \State Normalize measurement contributions
	    \For{$z_k \in z_{1:m}$}
	      \State $w^{(i), z_k} \leftarrow \frac{\hat{w}^{(i), z_k}}{\sum_{i' = 1}^J \hat{w}^{(i'), z_k} + \lambda_c c(z_k)}$
	    \EndFor
	    \State Update particle weight
	    \State $w^{(i)}_{+} \leftarrow w^{(i), \phi} + \sum_{z_k \in z_{1:m}} w^{(i), z_k}$
	  \EndFor
	\\
	\State \emph{Regional terms}
	  \State $\mu^{\phi}_{\Phi}(B) \leftarrow \sum_{x^{(i)} \in B} w^{(i), \phi}$
	  \For{$z_k \in z_{1:m}$}
	    \State $\mu^{z}_{\Phi}(B) \leftarrow  \sum_{x^{(i)} \in B} w^{(i), z_k}$
	  \EndFor
	\\
	\State \emph{Mean target number}
	  \State $\mu_{\Phi_{+}}(B) \simeq \mu^{\phi}_{\Phi}(B) + \sum_{z_k \in z_{1:m}} \mu^{z}_{\Phi}(B)$
	\\
	\State \emph{Variance in target number}	    
	  \State $var_{\Phi_{+}}(B) \simeq \mu^{\phi}_{\Phi}(B) + \sum_{z_k \in z_{1:m}} \mu^{z}_{\Phi}(B)\left(1 - \mu^{z}_{\Phi}(B)\right)$
      \end{algorithmic}
      \caption{PHD filter with variance: data update \cite{Vo_BN_2005} and information statistics\label{AlgPHD}}
    \end{algorithm}
  
  \bibliography{Bibliography.bib}
  \bibliographystyle{IEEEtran.bst}
  
\end{document}